%% file: main.tex
\documentclass[twocolumn, twocolappendix]{aastex63}

\usepackage{amsmath}
\usepackage{amssymb}
\usepackage{physics}
\usepackage{float}
\usepackage{booktabs}
\usepackage{verbatim}
\usepackage{natbib}
\usepackage{bbm}
\usepackage{tabularx}
\usepackage[]{footmisc}

\definecolor{mpl_blue}{HTML}{1F77B4}
\definecolor{mpl_orange}{HTML}{FF7F0E}
\definecolor{mpl_green}{HTML}{2CA02C}
\definecolor{mpl_red}{HTML}{D62728}
\definecolor{rbp}{HTML}{663399}

\hypersetup{backref, breaklinks, plainpages=false, colorlinks=true, anchorcolor=cyan, linkcolor=mpl_red, citecolor=cyan, urlcolor=cyan, bookmarks=false}
\usepackage[all]{hypcap}
\usepackage[caption=false]{subfig}
\usepackage{color}

\defcitealias{aaa+23_dataset}{NG15dataset}
\defcitealias{aaa+23}{NG15gwb}

\begin{document}

\title{The NANOGrav 15-year Data Set: Search for Anisotropy in the Gravitational-Wave Background}

\shorttitle{NANOGrav 15-year Anisotropic Gravitational-Wave Background}
\shortauthors{NANOGrav}

\input{authors15yranisotropy}

\begin{abstract}
    The North American Nanohertz Observatory for Gravitational Waves (NANOGrav) has reported evidence for the presence of an isotropic nanohertz gravitational wave background (GWB) in its 15~yr dataset. However, if the GWB is produced by a population of inspiraling supermassive black hole binary (SMBHB) systems, then the background is predicted to be anisotropic, depending on the distribution of these systems in the local Universe and the statistical properties of the SMBHB population. In this work, we search for anisotropy in the GWB using multiple methods and bases to describe the distribution of the GWB power on the sky. We do not find significant evidence of anisotropy, and place a Bayesian $95\%$ upper limit on the level of broadband anisotropy such that $(C_{l>0} / C_{l=0}) < 20\%$. We also derive conservative estimates on the anisotropy expected from a random distribution of SMBHB systems using astrophysical simulations conditioned on the isotropic GWB inferred in the 15-yr dataset, and show that this dataset has sufficient sensitivity to probe a large fraction of the predicted level of anisotropy. We end by highlighting the opportunities and challenges in searching for anisotropy in pulsar timing array data.
    
\end{abstract}

\keywords{Gravitational waves -- Black Holes -- Pulsars}

\section{Introduction} \label{sec:introduction}

    Using $67$ pulsars from its $15$-year data set \citep{aaa+23_dataset}, the North American Nanohertz Observatory for Gravitational Waves (NANOGrav; \citealp{NANOGrav,ransom+19}) has reported $\sim$$3-4\sigma$ evidence for a gravitational-wave background (GWB) in the frequency range $\sim$$2-30$~nHz \citep[][henceforth refered to as \citetalias{aaa+23}]{aaa+23}. This statistical significance is based on searching for the presence of a distinctive pattern of correlated timing deviations imprinted on otherwise-independent millisecond pulsars. These rotating neutron stars emit beams of radio waves that can intersect our line-of-sight every rotational period, registering pulses in radio telescopes. The regularity of their rotation, and the stability of their epoch-averaged radio pulse profiles, allow pulsars to be used as time-keepers with which we can build highly-accurate models of their rotation, position, proper motion, binary dynamics (where appropriate), and even the properties of the intervening interstellar medium through which the radio pulses must travel. Beyond being extraordinary objects in their own right, the timing stability of pulsars makes them excellent tools to study dynamic spacetime.
    
    A GW propagating between a pulsar and the Earth will cause a change in the proper length of the photon path, leading to deviations from expected pulse times of arrival (TOAs) that depend on the Earth-pulsar-GW geometry \citep{1975GReGr...6..439E,saz78,det79}. For an all-sky stochastic background of GWs, the net timing signature on a given pulsar may appear as a source of random noise with excess power on longer timescales \citep{phinney}. This is challenging to disentangle from other sources of noise such as intrinsic long-timescale pulsar rotational instabilities, or long-timescale variations in the properties of the interstellar medium \citep[see, e.g.,][and references therein]{aaa+23_noise}. Yet with Earth as the common end-point of all pulsar observations, the result of correlating the timing deviations between pairs of pulsars in a pulsar timing array (PTA) leads to an expected pattern that is akin to the overlap reduction function (ORF) of other GW detectors. For an isotropic GWB, this pattern is a quasi-quadrupolar signature that depends only on the angular separation of pulsars on the sky, known as the Hellings--Downs (HD) curve \citep{Hellings1983}. For anisotropic GWBs, this signature takes a different form \citep[e.g.,][]{ming_pta_anis}.

    The HD curve is used as a template when cross-correlating pulsar timing observations in detection statistics and for inferring the spectral properties of the GWB. It can be shown that in addition to arising due to an isotropic GWB, the HD curve is also the limiting case of binning pairwise correlations for an infinite number of pulsars due to an individual gravitational wave (GW) signal \citep{Cornish:2013aba,cosmic_variance_1}. Therefore, despite the fact that production-level PTA pipelines use unbinned correlation data for inference, the HD curve has been shown to be an effective template for initial detection of the GWB that may nevertheless be anisotropic \citep{Cornish:2013aba,cs16,bumpy_bkgrnd,2022ApJ...941..119B,cosmic_variance_1}. Indeed, NANOGrav's evidence for a GWB is based on the statistical significance of cross-correlations matching the HD curve versus the absence of any cross-correlations. Via a range of frequentist and Bayesian analyses that employed various simulations and data-augmentation techniques \citep{cs16,tlb+17, ppc1,ppc2}, NANOGrav constructed statistical background distributions for the significance of HD cross-correlations, resulting in a false-alarm probability of $\sim$$5\times10^{-5}-10^{-3}$ ($\sim$$3-4\sigma$).

    While the source of the GWB is not known for certain, one potential source is a population of inspiraling supermassive black-hole binaries (SMBHBs) with masses $10^8-10^{10}\,M_\odot$, whose superposition of quasi-monochromatic GW signals manifests as a stochastic background in our PTA. The inferred demographics and dynamics of such SMBHBs are studied in \citet{aaa+23_astro}. Other processes from the early Universe may contribute to the GWB \citep[see, e.g.,][and references therein]{aaa+23_cosmo}; however, an expected outcome of a SMBHB-dominated GWB is the presence of signal anisotropy, either through the clustering of host galaxies or Poisson fluctuations in GW source properties \citep{localSMBHBs2017}. GWB anisotropy is often described in terms of the angular power spectrum, wherein the GWB directional power map is decomposed on a spherical harmonic basis with associated coefficients, and then summarized with $C_l = \sum_{m=-l}^l |c_{lm}|^2/(2l+1)$. Estimates of GWB anisotropy from analytic and Monte-Carlo population studies produce angular power spectra where $(C_{l>0}/C_{l=0})\lesssim20\%$ \citep{ming_pta_anis,taylor_pta_anis,localSMBHBs2017}. Recent work by \citet{spk_model} also demonstrates the importance of high-frequency PTA sensitivity as a means of using anisotropy to probe details of the SMBHB population. 
    
    A variety of techniques have been developed to model and infer GWB anisotropy with PTA data, with broad similarities to how such searches are carried out in ground-based detectors \citep[e.g.,][]{1997PhRvD..56..545A,ballmer_radiometer,thrane_ligo_anis,RenziniContaldi2018,cg_anis_ligo,2023PhRvD.107d3016E}, and in plans for future space-borne detectors like LISA \citep[e.g.,][]{2022JCAP...11..009B,banagiri_blip,2001CQGra..18.4277C,ContaldiPieroni2020}. Differences arise mostly through the choice of basis on which to express the angular power distribution. \citet{taylor_pta_anis} developed the first Bayesian PTA pipeline for GWB anisotropy by expressing the angular power as a linear expansion of weighted spherical harmonics. The space of spherical-harmonic coefficients was bounded by a prior requiring that the GWB power be positive everywhere, which was assessed via rejection. The spherical-harmonic basis approach was fully generalized by \citet{Gair_pta_cmb_anis}, who computed an analytic form for ORF basis functions of any $\{lm\}$. Rather than imposing positivity via rejection, a more elegant approach was developed almost simultaneously for ground-based \citep{cg_anis_ligo}, space-borne \citep{banagiri_blip}, and PTA \citep{bumpy_bkgrnd} anisotropy searches through which the \textit{square root} of the GWB power was first expressed on a spherical-harmonic basis, thereby naturally imposing positive behavior on the GWB itself. Other techniques have been adapted from CMB analyses to map the polarization content of the GWB \citep{Gair_pta_cmb_anis,KatoSoda2016,hotinli,2022PhRvD.106b3004S,LiuNg2022}, although these have yet to be applied to real data. Finally, techniques using data-driven bases for anisotropy modeling show promise; by computing eigen-skies of the noise-weighted PTA response map, the GWB power distribution can be efficiently built from a compact number of basis terms \citep{cornish_eigenmaps,ali-hamoud_1,ali-hamoud-2}.
    
    The only dedicated PTA search for GWB anisotropy before now was performed by the European Pulsar Timing Array collaboration \citep{EPTA} using six high-quality pulsars \citep{epta_anisotropy}. With only $15$ distinct pulsar pair combinations, the prospects were limited for detecting anisotropy. They found that the Bayesian $95\%$ upper limit on the characteristic strain in higher spherical-harmonic multipoles---defined as $A_\mathrm{GWB}(C_l/4\pi)^{1/4}$---was $\lesssim40\%$ of the $l=0$ amplitude. However, this is almost entirely due to prior constraints enforcing a positive angular power distribution for the GWB, which limited the level of power in higher multipoles with respect to $l=0$. Almost by definition, data-informed constraints on GWB anisotropy require at least as many pulsars as are necessary for initial evidence of inter-pulsar correlations. It is only now that PTAs have reached this threshold, which motivates the search here. 

    This paper is organized as follows. In \S\ref{sec:methods} we discuss our methods for describing and searching for GWB anisotropy in the NANOGrav 15-year data set, including basis choices, and details of our Bayesian and frequentist pipelines. Our results are described in \S\ref{sec:results}, followed by a discussion in \S\ref{sec:discussion} that places these results in context using GWB anisotropy estimates from many realizations of SMBHB populations that were generated with NANOGrav's \texttt{holodeck} simulation software. We conclude and consider future prospects in \S\ref{sec:conclusion}.
    
\section{Methods} \label{sec:methods}

    We search for anisotropy in the GWB by using information from the full set of inter-pulsar correlations (i.e., auto-correlations and cross correlations) within a Bayesian analysis, as well as only cross-correlations within a frequentist framework.
    The search for anisotropy using all correlations is performed with the standard PTA Bayesian pipeline that is described in \citetalias{aaa+23}, with suitable modifications to account for anisotropy in the GWB. However, since this pipeline is slow for evaluating models with inter-pulsar correlations, we also perform a faster cross-correlation search using a frequentist approach based on the methods developed in \citet{pol_anisotropy}. In the following, we discuss the formalism for modeling GWB anisotropy in PTA data, followed by a description of our various analysis pipelines.

    \subsection{Overlap reduction function and the GWB power} \label{sec:ORF}

        For any PTA with $N_{\rm psr}$ pulsars, the total number of correlations is $N_{\rm tc} = N_{\rm psr} (N_{\rm psr} + 1) / 2$, of which there are $N_{\rm ac} = N_{\rm psr}$ auto correlations and $N_{\rm cc} = N_{\rm psr} (N_{\rm psr} - 1) / 2$ cross correlations. The angular dependence of these measured correlations on the distribution of GWB power is described by the ORF \citep{orf_citation,ming_pta_anis, taylor_pta_anis, Gair_pta_cmb_anis, bumpy_bkgrnd}, which for a Gaussian, stationary GWB can be written as,
        \begin{align} \label{eq:orf}
            \Gamma_{ab} \propto \int_{S^2} d^2\hat\Omega \,\,P(\hat\Omega) &\left[ \mathcal{F}^+(\hat{p}_a,\hat\Omega)\mathcal{F}^+(\hat{p}_b,\hat\Omega) \right. \nonumber\\
            &\left. +\, \mathcal{F}^\times(\hat{p}_a,\hat\Omega)\mathcal{F}^\times(\hat{p}_b,\hat\Omega) \right],
        \end{align}
        where $a,b$ index pulsars; $P(\hat\Omega)$ is the power of the GWB in direction $\hat\Omega$, normalized such that $\int_{S^2}d^2\hat\Omega \,\,P(\hat\Omega) = 1$; and $\mathcal{F}^A(\hat{p},\hat\Omega)$ is the antenna response of a pulsar in unit-vector direction $\hat{p}_a$ to each GW polarization $A\in[+,\times]$, defined such that
        \begin{equation}
            \displaystyle \mathcal{F}^A (\hat{p}, \hat{\Omega}) = \frac{1}{2} \frac{\hat{p}^i \hat{p}^j}{1 - \hat{\Omega} \cdot \hat{p}} e_{ij}^A (\hat{\Omega}),
            \label{eq:antenna_resp_def}
        \end{equation}
        where $e_{ij}^A(\hat{\Omega})$ are polarization basis tensors, and $(i, j)$ are spatial indices. Note that unlike ground- and space-based GW detectors, the GW-frequency dependence in the ORF can be factored out in the PTA regime, and we use the ORF to represent the angular dependence of the correlations \citep[e.g.,][]{romano_cornish_review}. If $P(\hat\Omega)=1,\,\, \forall\,\, \hat\Omega$, \autoref{eq:orf} is proportional to the HD curve. 
        
        The integral in \autoref{eq:orf} can be rewritten as a sum over equal-area pixels \citep{Gair_pta_cmb_anis, bumpy_bkgrnd} indexed by $k$,
        \begin{equation}
            \displaystyle \Gamma_{ab} \propto \sum_k P_k \left[ \mathcal{F}^+_{a,k}\mathcal{F}^+_{b,k} + \mathcal{F}^\times_{a,k}\mathcal{F}^\times_{b,k} \right]\Delta\hat\Omega_k.
            \label{eq:orf_discrete}
        \end{equation}

        To model GWB anisotropy and compute the ORF, we must choose an appropriate basis on the $2$-sphere to represent the GWB power. Here we model the GWB angular power dependence using a spherical-harmonic basis \citep{ming_pta_anis, Gair_pta_cmb_anis, taylor_pta_anis, epta_anisotropy} and a pixel basis \citep{cornish_eigenmaps}.

        \subsubsection{Radiometer pixel basis} \label{sec:pixel_basis}
            
            In the radiometer pixel basis \citep{ballmer_radiometer, mitra_radiometer}, the sky is divided into equal-area pixels using HEALPix \citep{healpix}:
            \begin{equation}
                \displaystyle P(\hat{\Omega}) = \sum_{\hat\Omega^{'}} P_{\hat{\Omega^{'}}} \delta^2(\hat{\Omega}, \hat\Omega^{'}),
                \label{eq:pixel_basis}
            \end{equation}
            such that the ORF for a given independently-modeled pixel is
            \begin{equation}
                \Gamma_{ab,\hat\Omega} \propto P_{\hat\Omega}\left[ \mathcal{F}^+_{a,\hat\Omega}\mathcal{F}^+_{b,\hat\Omega} + \mathcal{F}^\times_{a,\hat\Omega}\mathcal{F}^\times_{b,\hat\Omega} \right]\Delta\hat\Omega.
            \end{equation}
            The number of pixels on the sky is set by $N_{\rm pix} = 12 N_{\rm side}^2$, where $N_{\rm side}$ defines the tessellation of the HEALPix sky \citep{healpix}. For PTAs, the rule of thumb is to have $N_{\rm pix} \leq N_{cc}$ when counting pieces of information \citep{romano_cornish_review}. Given that $N_{\rm side}$ needs to be a power of 2 \citep{healpy}, this imposes a choice of $N_{\rm side} = 8$ for the 15 yr dataset with its 67 pulsars, resulting in an angular resolution of $\approx 7.3^{\circ}$.
            This basis is ideally suited for detecting widely separated point sources, since we assume that the power between any two neighbouring pixels is not correlated.
            
        \subsubsection{Spherical and square-root spherical harmonic basis} \label{sec:sph_basis}
        
        In the spherical-harmonic basis \citep{1997PhRvD..56..545A}, GWB power is written as a linear expansion over  the spherical-harmonic functions, which form an orthonormal basis on the $2$-sphere, such that
        \begin{equation}
            \displaystyle P(\hat{\Omega}) = \sum_{l = 0}^{\infty} \sum_{m = -l}^{l} c_{lm} Y_{lm}(\hat\Omega)
            \label{eq:sph_basis}
        \end{equation}
        where $Y_{lm}$ are the real valued spherical harmonics. Without prior restrictions or model regularization on the coefficients $c_{lm}$, the linear spherical-harmonic basis allows the GWB power to assume negative values, which is an unphysical model of the GWB. We can address this problem by instead modeling the square-root of the GWB power, $P(\hat{\Omega})^{1/2}$, rather than the power itself. This technique was introduced in a Bayesian context in \citet{cg_anis_ligo} for LIGO, \citet{banagiri_blip} for LISA and \citet{bumpy_bkgrnd} for PTAs, while \citet{pol_anisotropy} applied this method in a frequentist context for PTAs. The square-root of the power can be decomposed onto the spherical harmonic basis,
        \begin{equation}
            \displaystyle P(\hat{\Omega}) = [P(\hat{\Omega})^{1/2}]^2 = \left[ \sum_{L=0}^{\infty} \sum_{M=-L}^{L} b_{LM} Y_{LM}(\hat{\Omega})\right]^2,
            \label{eq:sqrt_sph_basis}
        \end{equation}
        where $Y_{LM}$ are the real valued spherical harmonics and $b_{LM}$ are the search coefficients. \citet{banagiri_blip} showed that the search coefficients in the square-root spherical-harmonic basis can be related to the coefficients in the linear basis via
            \begin{equation}
                \displaystyle c_{lm} = \sum_{LM} \sum_{L^{\prime} M^{\prime}} b_{LM} b_{L^{\prime} M^{\prime}} \beta_{lm}^{LM, L^{\prime} M^{\prime}},
                \label{eq:sqrt_to_sph}
            \end{equation}
            where $\beta_{lm}^{LM, L^{\prime} M^{\prime}}$ is defined as
            \begin{equation}
                \displaystyle \beta_{lm}^{LM, L^{\prime} M^{\prime}} = \sqrt{ \frac{(2L + 1) (2L^{\prime} + 1)}{4 \pi (2l + 1)}} C^{lm}_{LM, L^{\prime} M^{\prime}} C^{l0}_{L0, L^{\prime} 0},
                \label{eq:cg_coeff}
            \end{equation}
            with $C^{lm}_{LM, L^{\prime} M^{\prime}}$ being Clebsch-Gordon coefficients. This approach imposes control on the spherical-harmonic coefficients to inhibit the proposal of GWB power distributions with negative regions.
            
            We quantify our results from the spherical harmonic basis in terms of $C_{l}$, which is the squared angular power in each multipole mode $l$,
            \begin{equation}
                \displaystyle C_l = \frac{1}{2l + 1} \sum_{m = -l}^{l} \left| c_{lm} \right|^2.
                \label{eq:Cl}
            \end{equation}
            $C_{l}$ is thus a measure of the amplitude of the statistical fluctuations in the angular power of the GWB at scales $\theta = 180^{\circ} / l$. An isotropic GWB will only have power in the $l = 0$ multipole (typically referred to as the monopole), while an anisotropic GWB will have power at higher-$l$ multipoles.
            As shown in \citet{pen_boyle_pta_resolution}, the diffraction limit defines the highest multipole, $l_{\rm max}$, that can be probed in an anisotropic search, which for PTAs scales as $l_{\rm max} \sim \sqrt{N_{\rm psr}}$ \citep{romano_cornish_review}, and is $l_{\rm max} \approx 8$ for the NANOGrav 15 yr dataset, giving a maximum angular resolution of $\theta = 22.5^{\circ}$, which is approximately three times larger than the resolution of the radiometer pixel basis. Thus, the spherical harmonic and radiometer pixel bases are probing anistropies on large and small angular scales respectively.
        
    \subsection{Bayesian analysis pipeline} \label{sec:bayesian}
    
        The Bayesian pipeline is designed to use the full correlation data available to PTAs, i.e., both the spatial auto- and cross-correlations between pulsars in the array \citep{bumpy_bkgrnd}. Assuming an unpolarized, wide-sense stationary Gaussian GWB, the ORF from \autoref{eq:orf_discrete} can be rewritten in matrix form \citep{bumpy_bkgrnd},
        \begin{equation}
            \displaystyle \mathbf{\Gamma}_{ab} = [\mathbf{F} \mathbf{P} \mathbf{F}^{T}]_{ab},
            \label{eq:orf_bayesian}
        \end{equation}
        where $\mathbf{P}$ is a diagonal matrix of size $2N_\mathrm{pix}$ describing the GWB power in each polarization for each pixel, and $\mathbf{F}$ represents the PTA signal response matrix of size $N_\mathrm{psr}\times 2N_\mathrm{pix}$, which can be split into Earth and pulsar term contributions as
        \begin{align}
            \displaystyle & F^E_{\{a,k,A\}} = \sqrt{\frac{3}{2}} \frac{\mathcal{F}^A(\hat{p}_a, \hat{\Omega}_k)}{N_{\rm pix}} \nonumber \\
            & F^P_{\{a,k,A\}} = F^E_{\{a,k,A\}} e^{-2\pi i f L_a (1 - \hat{\Omega}_k \cdot \hat{p}_a) / c},
            \label{eq:signal_response_bayesian}
        \end{align}
        where $f$ is the GW frequency, $L_a$ is the distance to pulsar $a$, sky pixels are indexed by $k$, and GW polarization is labeled by $A\in[+,\times]$. Since the distance to pulsars is $\sim$parsecs or more, the Earth and pulsar term components are numerically orthogonal due to the rapid oscillation of the pulsar term across the sky  \citep{MingarelliSidery2014,MM18, bumpy_bkgrnd}. Thus, the pulsar term contribution to the ORF is effectively diagonal, such that
        \begin{equation}
            \mathbf{\Gamma}_{ab} = (1 + \delta_{ab}) [\mathbf{F}^E \mathbf{P} \mathbf{F}^{E,T}]_{ab},
            \label{eq:bayesian_orf_full}
        \end{equation}
        where the GWB power matrix $\mathbf{P}$ can be expressed in any convenient basis, since the PTA signal response matrices sandwiching it are performing the necessary sky integral through a numerical sum over pixels.

        The Bayesian pipeline is constructed identically to the one described in \citetalias{aaa+23}. Given an ORF $\Gamma_{ab}$, the GWB cross-power spectrum can be written as $\Phi_{ab,j} = \Gamma_{ab} \Phi_j$, where $\Phi_{j}$ represents the usual auto-power spectrum. We consider two different ways to parametrize $\Phi_j$: $(1)$ referred to as the ``free spectrum'' analysis, we model all $\Phi_j$ components independently and simultaneously; and $(2)$ we assume a power-law spectral template across frequencies, in which case
        \begin{equation}
            \Phi_j = \frac{A^2}{12 \pi^2} \frac{1}{T} \left( \frac{f_j}{f_{\rm ref}} \right)^{-\gamma} f^{-3}_{\rm ref},
            \label{eq:power-law_template}
        \end{equation}
        where $A$ is the characteristic strain at a reference frequency of $f_{\rm ref} = 1 {\rm yr}^{-1}$, $\gamma$ is the spectral index of the GWB's power spectral density, and $T$ is the total timing baseline of the dataset. Approach-$(1)$ allows us to search for frequency-resolved anisotropy (i.e., we can measure anisotropy independently and simultaneously at all frequencies), while approach-$(2)$ searches for broadband anisotropy modeled on a GWB power-law spectrum. However, frequency-resolved anisotropy searches are computationally expensive given the large parameter space that must be explored, while searches that assume a power-law template for the GWB are more tractable. 
        
        The construction of the likelihood and priors, and the sampling techniques are identical to those used in \citetalias{aaa+23}. 
        For the linear spherical harmonic basis, we set uniform priors with boundaries $[-5, 5]$ on the spherical harmonic coefficients, $c_{lm}$, and fix $c_{00} = \sqrt{4\pi}$ such that the variation in the monopole is modeled by the amplitude of the GWB. For the square-root spherical harmonic basis, we fix $b_{00} = 1$ to break scale and parity symmetries \citep{banagiri_blip}. Additionally, in this basis, the $\{ b_{lm} \}$ parameters are complex-valued for $m \neq 0$, and we set the priors on the two degrees of freedom, the amplitude, $\left| b_{lm} \right|$, and phase, $\phi_{lm}$, to be uniform between $[0, 50]$ and $[0, 2\pi]$ respectively. The $b_{l0}$ parameters are real-valued \citep{banagiri_blip}, and we set uniform priors between $[-50, 50]$ for these parameters.
        
        We measure the evidence for the presence of anisotropy by calculating the odds ratio between an anisotropic and isotropic GWB models. Given the computational cost of the frequency-resolved anisotropy search, we measure the evidence for the presence of anisotropy in this analysis by calculating the Hellinger distance metric \citep{hellinger_dist} between the prior and posterior distributions of the angular power spectrum at each frequency. For two discrete probability distributions, $P = (p_1, \ldots, p_k)$ and $Q = (q_1, \ldots, q_k)$, the Hellinger distance is defined as \citep{hellinger_dist},
        \begin{equation}
            \displaystyle H(P,Q) = \frac{1}{\sqrt{2}}\sqrt{\sum_{i=1}^{k} \left( \sqrt{p_i} - \sqrt{q_i} \right)^2}.
            \label{eq:hellinger_dist}
        \end{equation}
        This is a bounded metric, $0 \leq H(P, Q) \leq 1$, such that a distance of $0$ and $1$ imply that $P$ and $Q$ are completely identical and distinct respectively.
        
    \subsection{Frequentist analysis pipeline} \label{sec:frequentist}

        The frequentist pipeline is based on using the pulsar cross-correlations as data, as described in \citet{pol_anisotropy}. The cross-correlations between two pulsars $a$ and $b$, $\rho_{ab}$, and their uncertainties, $\sigma_{ab}$, are defined as \citep{NG5yr_OS, siemens_scaling_laws, optimal_statistic_chamberlin, noise_marg_os_vigeland},
        \begin{align} \label{eq:cross_corr}
            \displaystyle \rho_{ab} &= \frac{\delta\textbf{t}^T_{a} \textbf{N}^{-1}_{a} \hat{\textbf{S}}_{ab} \textbf{N}^{-1}_{b} \delta\textbf{t}^T_{b}}{\textrm{tr}\left[ \textbf{N}^{-1}_{a} \hat{\textbf{S}}_{ab} \textbf{N}^{-1}_{b} \hat{\textbf{S}}_{ba} \right]}, \nonumber\\
            \displaystyle \sigma_{ab} &= \left( \textrm{tr}\left[ \textbf{N}^{-1}_{a} \hat{\textbf{S}}_{ab} \textbf{N}^{-1}_{b} \hat{\textbf{S}}_{ba} \right] \right)^{-1/2},
        \end{align}
        where $\delta\mathbf{t}_a$ is a vector of timing residuals for pulsar $a$, $\textbf{N}_a = \langle \delta \mathbf{t}_a \delta \mathbf{t}_a^T\rangle$ is the measured autocovariance matrix of pulsar $a$, and $\hat{\textbf{S}}_{ab}\equiv \langle \delta \mathbf{t}_a \delta \mathbf{t}_b^T\rangle / \Gamma_{ab}A^2$ is the template-scaled covariance matrix between pulsar $a$ and $b$. We use the formalism of \citet{noise_marg_os_vigeland} to calculate the ``noise marginalized'' optimal statistic (NMOS) cross-correlations and their uncertainties over multiple random draws from the posterior samples of a common uncorrelated red-noise Bayesian analysis (see \citetalias{aaa+23} for more details). This allows us to produce frequentist results that also marginalize over the intrinsic pulsar noise, similar to the Bayesian analyses. 
        
        PTA cross-correlation data can be modeled with the ORF from \autoref{eq:orf_discrete}, which can be further simplified into a general matrix form as $\mathbf{\Gamma} = \mathbf{R} \mathbf{P} \label{eq:orf_frequentist}$,
        where $\mathbf{\Gamma}$ is an $N_{\rm cc}$ vector of ORF values for all distinct pulsar pairs, $\mathbf{P}$ is a vector describing the GWB power, and $\mathbf{R}$ is a PTA overlap response matrix given by
        \begin{equation}
            \displaystyle R_{ab,k} = \frac{3}{2N_{\rm pix}} \left[  \mathcal{F}^+_{a,k}\mathcal{F}^+_{b,k} + \mathcal{F}^\times_{a,k}\mathcal{F}^\times_{b,k}  \right],
            \label{eq:overlap_response_matrix_frequentist}
        \end{equation}
        where the normalization is chosen so that the ORF matches the HD values in the case of an isotropic GWB. We use the \textsc{maps} software package \citep{pol_anisotropy}, which can model the GWB power in both the (normal and square-root) spherical-harmonic and pixel bases. As shown in \citet{pol_anisotropy}, the cross-correlation likelihood can be written as
        \begin{equation}
            \displaystyle p(\boldsymbol{\rho} | \mathbf{P}) =  \frac{\textrm{exp} \left[ -\frac{1}{2} (\boldsymbol{\rho} - \mathbf{R} \mathbf{P})^T \, \mathbf{\Sigma}^{-1} \, (\boldsymbol{\rho} - \mathbf{R} \mathbf{P}) \right]}{\sqrt{\mathrm{det}(2\pi\mathbf{\Sigma})}},
            \label{eq:frequentist_lkl}
        \end{equation}
        where $\mathbf{\Sigma}$ is the diagonal covariance matrix of cross-correlation uncertainties, with shape $N_{\rm cc} \times N_{\rm cc}$.

        For the radiometer pixel and (linear) spherical-harmonic basis, the problem is linear in the regression coefficients (i.e., pixel amplitude and spherical harmonic coefficients). So the maximum likelihood solution can be derived analytically \citep{thrane_ligo_anis, romano_cornish_review, ivezic_book}:
        \begin{equation}
            \displaystyle \mathbf{\hat{P}} = \mathbf{M}^{-1} \mathbf{X},
            \label{eq:max_lkl}
        \end{equation}
        where $\mathbf{M} = \mathbf{R}^T \mathbf{\Sigma}^{-1} \mathbf{R}$ is the Fisher information matrix, with uncertainties on the regression parameters given by the diagonal elements of $\mathbf{M}^{-1}$, and $\mathbf{X} = \mathbf{R}^T \mathbf{\Sigma}^{-1} \boldsymbol{\rho}$ is the ``dirty map", an inverse-noise weighted representation of the total power on the sky as seen through the response of the pulsars in the PTA. Note that for the radiometer pixel basis, since the individual pixels are modeled independently, the inverse of the full Fisher matrix in \autoref{eq:max_lkl} is replaced by the inverse of the diagonal elements of the Fisher matrix to obtain an estimate of the amplitude in any given pixel \citep{romano_cornish_review}. For the square-root spherical harmonic basis, since the problem is non-linear in the regression coefficients (i.e., the $b_{lm}$ parameters), the maximum likelihood solution is derived using numerical optimisation techniques. As described in \citet{pol_anisotropy}, we use the \textsc{lmfit} \citep{lmfit} Python package with Levenberg-Marquardt optimisation \citep{levenberg1944method, marquardt1963algorithm} to calculate the maximum likelihood solution. 

        For the spherical harmonic basis, we define three types of signal-to-noise (S/N) ratios through the maximum likelihood ratio \citep{pol_anisotropy}: (i) the total S/N, defined as the maximum likelihood ratio between an anisotropic model and noise; (ii) isotropic S/N, defined as the maximum likelihood ratio between an isotropic model and noise; and (iii) anisotropic S/N, defined as the maximum likelihood ratio between an anisotropic model and an isotropic model. Together, these S/N ratios provide a complete description of the correlations that might be present in the data. To interpret the S/N values, we calibrate them against a null distribution that is constructed using the measured uncertainties for the pairwise cross-correlations. Since the null hypothesis when searching for anisotropy is isotropy, we construct the null hypothesis by generating draws for each pulsar pair from a Gaussian distribution whose mean is the theoretical HD value and standard deviation is the cross correlation uncertainty measured from the real dataset. For each of these ``realizations'' of the null distribution, we calculate the S/N ratios and calibrate the significance of the S/N values measured with the real dataset through $p$-values, where a $p$-value $p < 3\times10^{-3}$ (corresponding to a 3$\sigma$ Gaussian-equivalent threshold) would imply a significant detection of anisotropy. We also use these null distributions to define the ``decision threshold'', $C_{l}^{\rm th}$, for each spherical harmonic multipole, such that if the measured angular power at any multipole is above this decision threshold, that would imply the measured angular power is inconsistent with the null hypothesis at the ~3$\sigma$ level \citep{pol_anisotropy}. The decision threshold allows us to search for the presence of anisotropy at a single multipole, while the S/N ratios represent holistic evidence for the presence of anisotropy in the data.

        For the radiometer pixel basis, we define our detection statistic as the ratio of the power, $P_{\hat\Omega}$, measured in each pixel to the uncertainty, $\sigma_{P_{\hat\Omega}}$, on that measurement, i.e., $P_{\hat\Omega} / \sigma_{P_{\hat\Omega}}$. For each of the realizations of the null hypothesis described above, we calculate the corresponding sky maps and uncertainties, compute the detection statistic for each pixel, and construct the null distribution for each individual pixel across realizations of the null hypothesis. We use this null distribution per pixel in conjunction with dividing our $p$-value threshold of $3\times 10^{-3}$ ($\sim$3$\sigma$) by a trials factor, $n = N_{\rm pix}$, to calibrate the significance of the detection statistic measured in the real data.
    
\section{Results} \label{sec:results}

    In \citetalias{aaa+23}, the low-frequency GWB signal is described using the lowest fourteen frequency bins when using the full set of correlations. However, most of the support for HD correlations in the data is concentrated in the lowest five frequency bins, with the higher frequency bins showing evidence for the auto-correlations. As a result, we use the lowest five frequency bins to model the anisotropy in the GWB, but perform a few analyses using fourteen frequency bins and find no significant difference from our results using the lowest five bins.

    In the Bayesian analysis, we model the GWB as a power-law with both a fixed $\gamma = 13/3$ \citep{phinney} and varied spectral index.
    We also model the anisotropy at each of the lowest five Fourier frequency bins simultaneously, as different SMBHB systems will contribute to different frequency bins resulting in unique anisotropy signatures at different frequencies, which might not be detectable under a power-law template for the GWB.   
    The spectral template of the GWB in the frequentist analyses is limited to a power-law template, again using just the lowest five frequency bins. Given the results from \citetalias{aaa+23}, we search for anisotropy at spectral indices of $\gamma = 3.2$, corresponding to the maximum a-posteriori value, and $\gamma = 13/3$. 

    \begin{figure}[htb]
        \centering
        \includegraphics[width = \columnwidth]{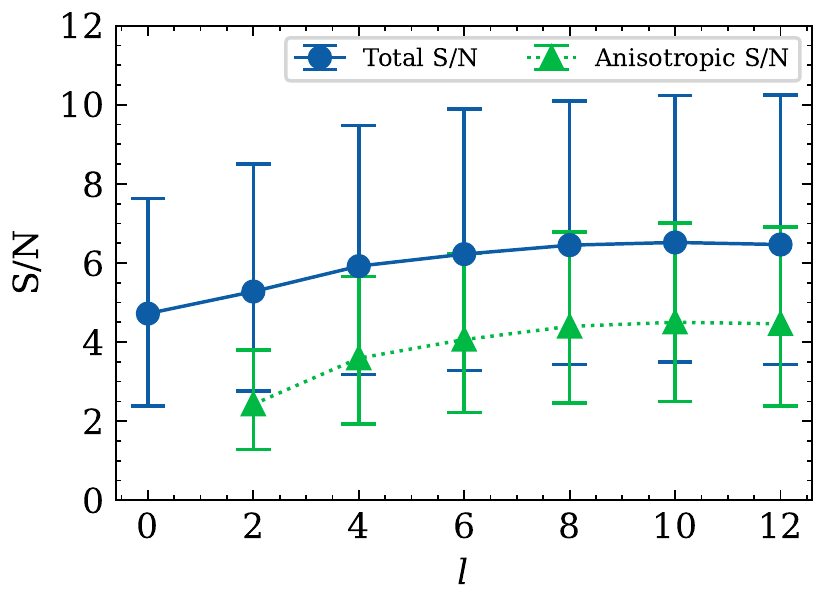}
        \caption{The total and anisotropic signal-to-noise (S/N) ratio derived using the frequentist square-root spherical harmonic basis described in Sec.~\ref{sec:frequentist}. Both the total and anisotropic S/N start to saturate at an $l_{\rm max} = 6$, which we choose for all the analyses presented in this work.}
        \label{fig:ml_sn}
    \end{figure}
    
    As described in Sec.~\ref{sec:pixel_basis} and \ref{sec:sph_basis}, the diffraction limit defines an optimal choice of $N_{\rm side} = 8$ and $l_{\rm max} = 8$ for the pixel and spherical harmonic bases respectively. \citet{higher_lmax_limit} showed that searches at multipoles higher than $l_{\rm max}$ can be feasible, though this results in a reduction in the overall S/N of any anisotropic signal that might be present in the data. 
    To test whether the choice of $l_{\rm max}$ defined by the diffraction limit is supported by the data, we calculate the maximum likelihood S/N values as a function of $l_{\rm max}$. As shown in \autoref{fig:ml_sn}, we see that the total and anisotropic S/N ratios start to saturate at $l_{\rm max} = 6$, slightly lower than the diffraction-limit implied $l_{\rm max} = 8$. To prevent over-fitting the data, we set $l_{\rm max} = 6$ for all of our analyses.

    \subsection{Detector antenna response}

        The directional response of any single pulsar, $a$, in the PTA to the presence of a (anisotropic) GWB is quantified through the antenna response pattern in \autoref{eq:antenna_resp_def}, while the response of the correlations between a pair of pulsars, $a, b$, is quantified by the term in the parentheses in \autoref{eq:orf} and \autoref{eq:orf_discrete} \citep{romano_cornish_review}. However, given that no two pulsars in the PTA are identical, it is important to weight the response of each pulsar pair by the corresponding uncertainty on the measured cross correlations between that pair of pulsars. 

        \begin{figure}
            \centering
            \includegraphics[width = \columnwidth]{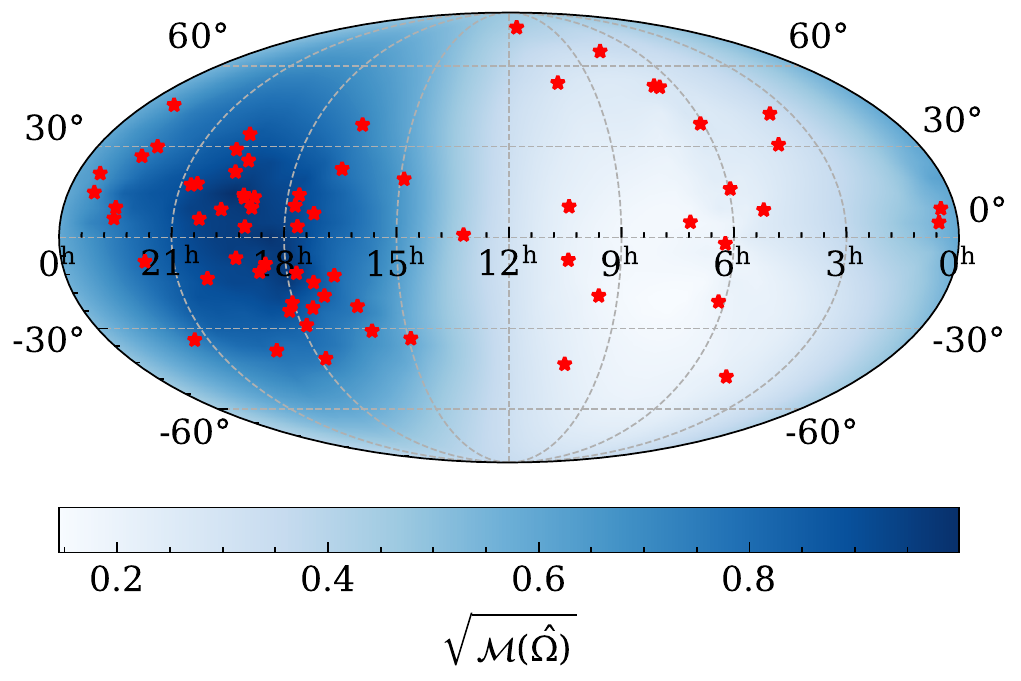}
            \caption{The normalized square-root of the diagonal elements of the Fisher matrix over 5000 draws from the NMOS cross correlation uncertainties, representing the relative sensitivity of the PTA to anisotropy in different directions on the sky. As expected, the PTA has highest sensitivity in the part of the sky that has the highest density of pulsars.}
            \label{fig:ant_response}
        \end{figure}
        
        The directional sensitivity of the PTA can thus be gauged through the diagonal elements of the Fisher matrix, $\mathbf{M}$, introduced in Sec.~\ref{sec:frequentist}. \autoref{fig:ant_response} shows the median of $\sqrt{\mathcal{M}(\hat{\Omega})}$, the normalized square-root of the diagonal elements of the Fisher matrix, across 5000 draws from the NMOS. 
        Since $\sqrt{\mathbf{M}^{-1}}$ represents the uncertainty on the amplitude measured in each pixel in the radiometer pixel basis, this map represents the relative sensitivity of the NANOGrav 15 yr dataset to different directions on the sky.
        As we can see, the detector has the highest sensitivity where it has the highest density of pulsars. Note that the cross-correlation uncertainties used in calculating this map are derived from using a power-law template for the GWB, and the detector antenna response may be slightly different at different frequencies.
        
    \subsection{Spherical harmonic basis} \label{subsec:sph_basis}
        
        \begin{figure}[htb]
            \centering
            \subfloat{\includegraphics[width = \columnwidth]{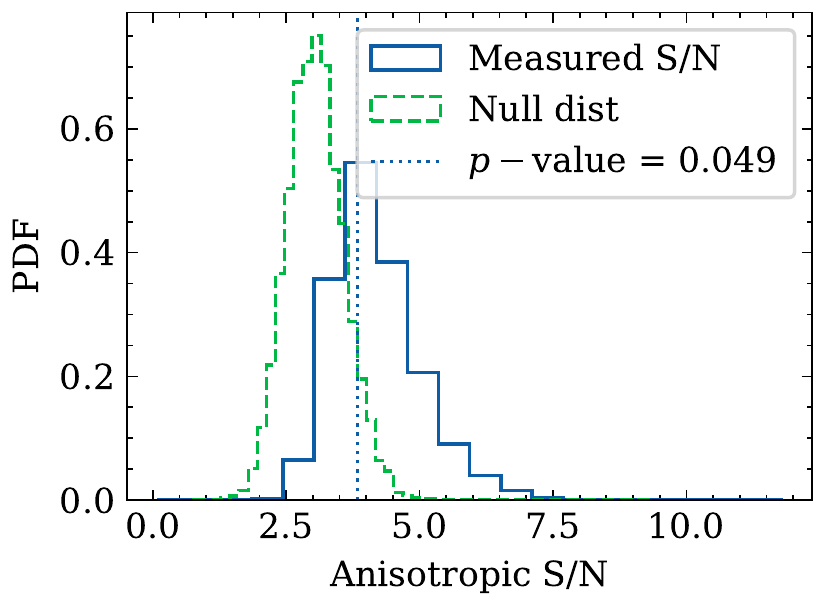}}
            \caption{The noise-marginalized distribution of the anisotropic S/N is shown along with the anisotropic S/N derived under the null hypothesis of an isotropic GWB (Sec.~\ref{sec:frequentist}). The mode of the measured anisotropic S/N corresponds to a $p$-value of $p = 0.05$, which is greater than our $p < 3\times10^{-3}$ threshold (Sec.~\ref{sec:frequentist}), implying that we do not have a significant detection of anisotropy in this dataset.}
            \label{fig:anis_sn}
        \end{figure}
    
        \begin{figure}[htb]
            \centering
            \includegraphics[width = \columnwidth]{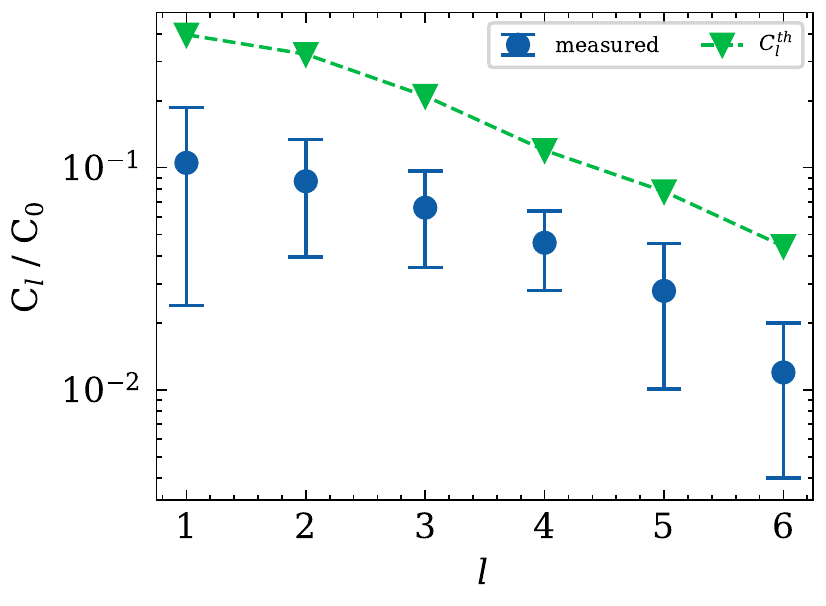}
            \caption{The angular power spectrum and decision threshold (Sec.~\ref{sec:frequentist}) measured using the frequentist square-root spherical harmonic basis. The measured power does not rise above the decision threshold for any multipole, implying that the data are consistent with isotropic GWB.}
            \label{fig:os_cl}
        \end{figure}
        
        We show the distribution of the measured anisotropic S/N measured from 5000 draws from the NMOS in \autoref{fig:anis_sn}, along with the distribution for the S/N under the null hypothesis of an isotropic GWB. We measure an anisotropic S/N of $\approx$4, which corresponds to a significance at the $p = 0.05$ level.  
        Thus, while there is some evidence for the presence of anisotropy in the 15 yr dataset, it does not yet rise to the level of a ``significant'' detection, i.e. $p > 3\times10^{-3}$ (Sec.~\ref{sec:frequentist}).  
        We also measure the angular power at each multipole, as shown in \autoref{fig:os_cl} along with the decision threshold (see Sec.~\ref{sec:frequentist}). As the power in any of the multipoles does not rise above the decision threshold, the data are consistent with isotropy at all spherical harmonic multipoles. 
        
        \begin{figure*}[htb]
            \centering
            \includegraphics[width = \textwidth]{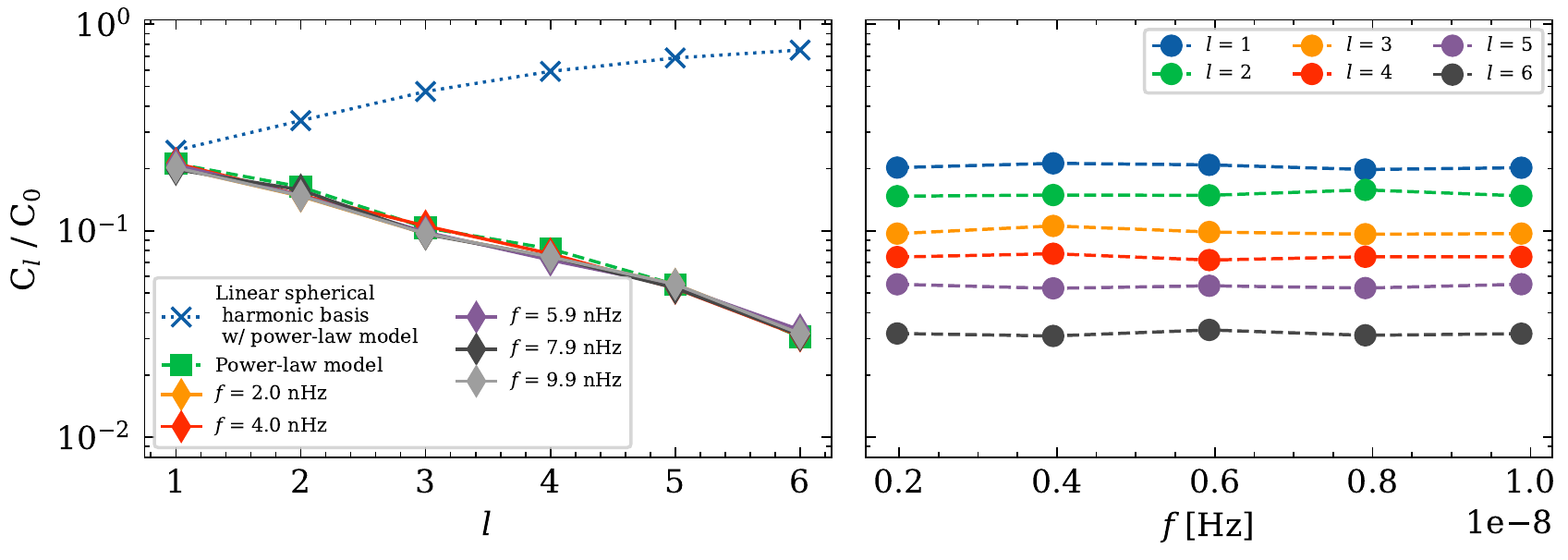}
            \caption{\textit{Left:} The 95\% upper limits on the angular power spectrum derived from the Bayesian linear (dotted line with ``x'' marker) and square-root (dashed line with square marker) spherical harmonic basis using a lowest-five-frequency-bin power-law template for the anisotropic GWB while varying the spectral index. For the latter, we also show the 95\% upper limits on the angular spectrum when searching for anisotropy in the lowest five bins simultaneously. For the sqaure-root spherical harmonic basis, the evolution of the upper limits as a function of spherical harmonic multipoles reflects the constraint from the prior condition that the power be positive on the sky as imposed by the square-root spherical harmonic basis. \textit{Right:} The 95\% upper limits at different spherical harmonic multipoles as a function of frequency for the square-root spherical harmonic basis. The lack of frequency dependence in the upper limits implies that the sensitivity to anisotropy at each frequency is defined by the sensitivity of the overall PTA to the GWB at each of those frequencies.}
            \label{fig:bayesian_cl}
        \end{figure*}

        \begin{figure}
            \centering
            \subfloat{\includegraphics[width = 0.5\textwidth]{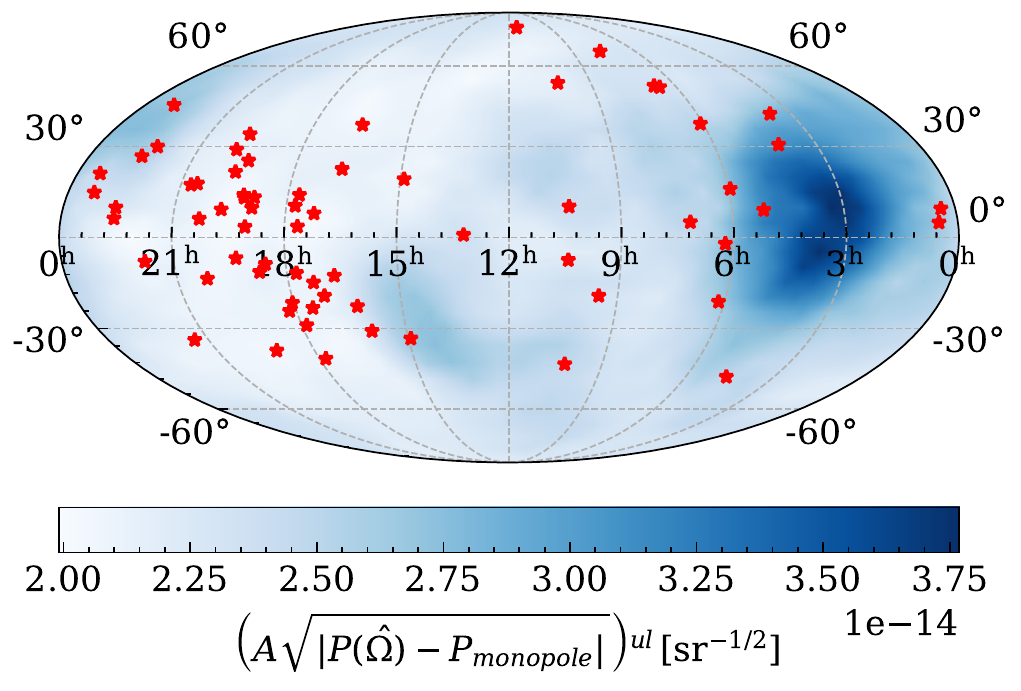}}
            \caption{Reconstructed sky map from the Bayesian square-root spherical harmonic basis showing the 95\% upper limit on deviations away from the isotropic component of the GWB, where the red stars represent the position of pulsars in the PTA. The power is represented in terms of the characteristic strain, referenced to a frequency of $f_{\rm ref} = 1 {\rm yr}^{-1}$, per steradian$^{1/2}$, marginalizing over the spectral index of the GWB. The odds ratio in favor of anisotropy over isotropy for the GWB $\approx$2, implying no significant detection of anisotropy in this dataset.}
            \label{fig:pl_maps}
        \end{figure}

\begin{figure*}[htb]
            \centering
            \includegraphics[width = \textwidth]{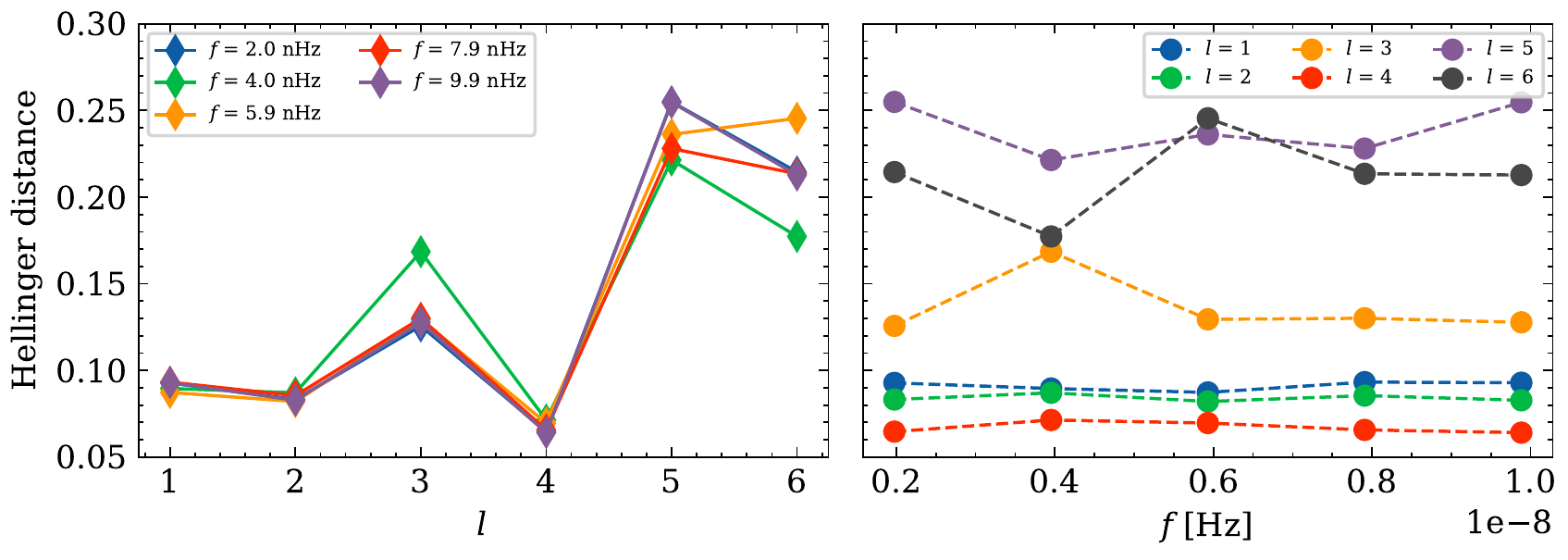}
            \caption{The Hellinger distance between posterior and prior distributions of $C_l / C_0$ for the frequency resolved anisotropy analysis. A Hellinger distance of 0 and 1 imply that the two distributions are identical and distinct respectively.  \textit{Left:} The Hellinger distance for different frequencies as a function of spherical harmonic multipoles. \textit{Right:} The Hellinger distance for different spherical harmonic multipoles as a function of frequency. This plot shows that there is no significant difference between the posterior and prior distribution of the angular power for any of the lowest five frequencies, implying that the data are consistent with isotropy in these bins.}
            \label{fig:bayesian_bin2}
        \end{figure*}

\begin{figure*}
            \centering
            \includegraphics{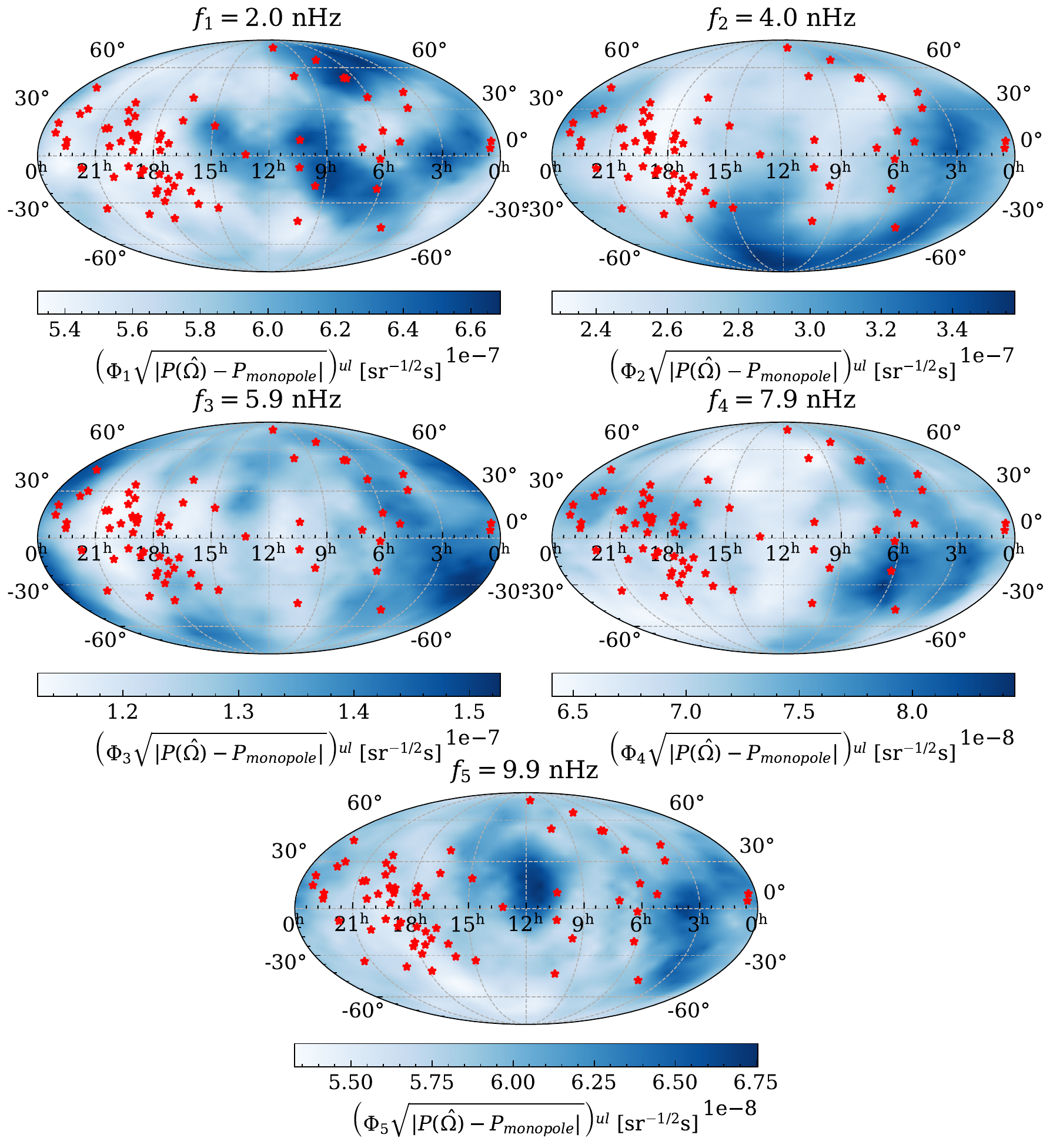}
            \caption{Reconstructed frequency resolved sky maps from the Bayesian square-root spherical harmonic basis showing the 95\% upper limit on the deviation away from the isotropic component of the GWB. The red stars represent the position of the pulsars in the PTA. The units for the power in these maps are seconds per steradian$^{1/2}$ and represent the excess timing delay, adopted for ease of comparison to Fig. 1(a) in \citetalias{aaa+23}.}
            \label{fig:fs_pow_map}
        \end{figure*}

        The search for anisotropy using our Bayesian pipeline produces results that are consistent with the frequentist results described above. We find an odds ratio of $\approx$2 in favour of an anisotropic model for the GWB using the square-root spherical harmonic basis, consistent with the non-detection in the frequentist analysis.
        Given the lack of detection, we plot the 95\% credible region on the angular power spectrum calculated using both the linear and square-root spherical harmonic bases in \autoref{fig:bayesian_cl} and show the reconstructed 95\% upper-limits on deviations away from isotropy in \autoref{fig:pl_maps}. 
        We also search for anisotropy simultaneously in each of the lowest five frequency bins of the detector, and find that our analysis returns the priors, as shown using the Hellinger distance metric in \autoref{fig:bayesian_bin2} implying no significant detection of anisotropy. Consequently, we show the 95\% upper limits for the angular power spectrum at each of these frequencies in \autoref{fig:bayesian_cl}, 
        and show the reconstructed 95\% upper-limits on deviations away from isotropy for these five frequencies in \autoref{fig:fs_pow_map}.
        
    \subsection{Radiometer pixel basis} \label{subsec:radio_pixel}

\begin{figure}
            \centering
            \subfloat{\includegraphics[width = 0.5\textwidth]{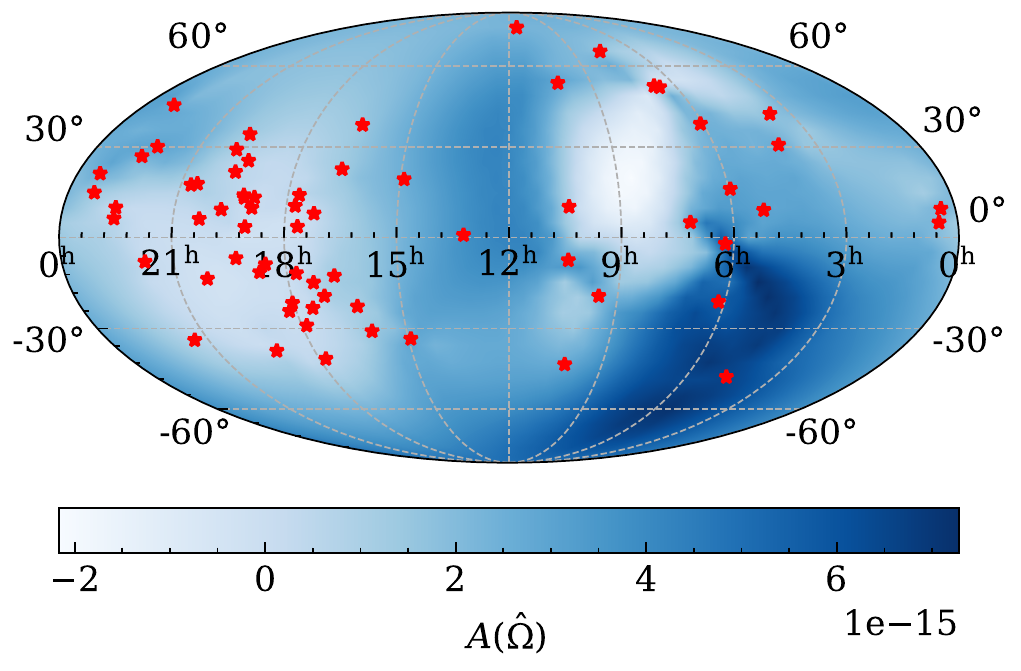}}
            \hfill
            \subfloat{\includegraphics[width = 0.5\textwidth]{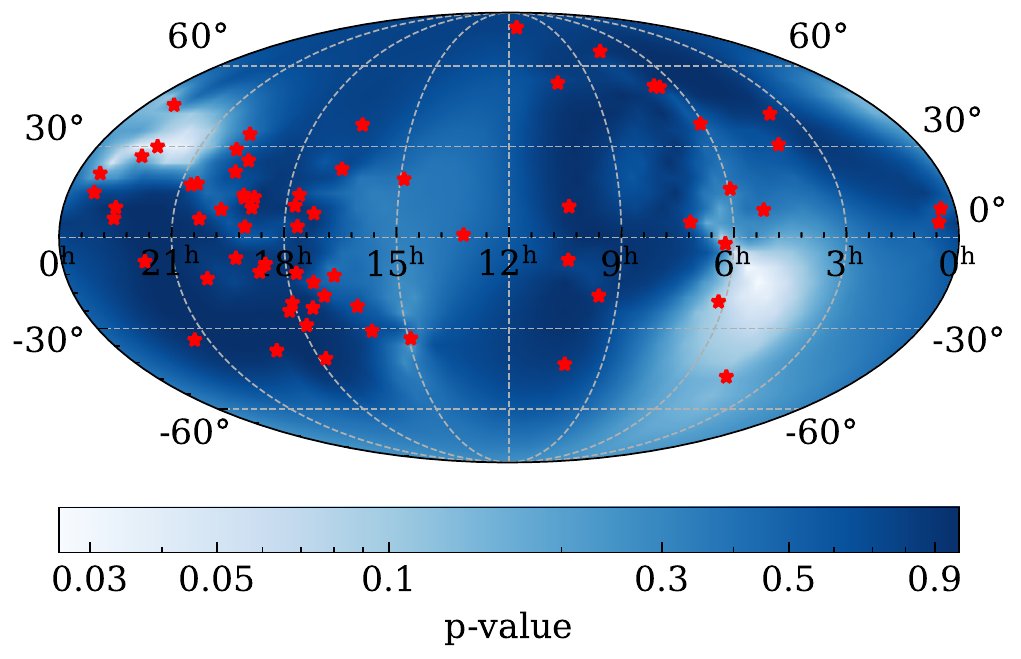}}
            \caption{\textit{Top:} Power measured at each pixel using the frequentist radiometer pixel basis, in units of characteristic strain. This map was constructed with $N_{\rm side} = 8$, resulting in 768 pixels on the sky. The negative recovered power is a consequence of this analysis not imposing the condition that the power be positive across all sky. \textit{Bottom:} The $p$-value corresponding to the detection statistic for each pixel shown in the top panel. All of the measured $p$-values imply that the data are consistent with isotropy.}
            \label{fig:freq_radio_pixel}
        \end{figure}

        \autoref{fig:freq_radio_pixel} shows the median of the noise-marginalized radiometer pixel map, along with the $p$-values corresponding to each pixel, calculated as described in Sec.~\ref{sec:frequentist}, and show that we do not detect significant power in any single pixel. Note that the frequentist analysis for this basis does not impose the condition that the power be positive across all sky, resulting in negative power in some parts of the sky.
        
\begin{figure}
            \centering
            \subfloat{\includegraphics[width = 0.5\textwidth]{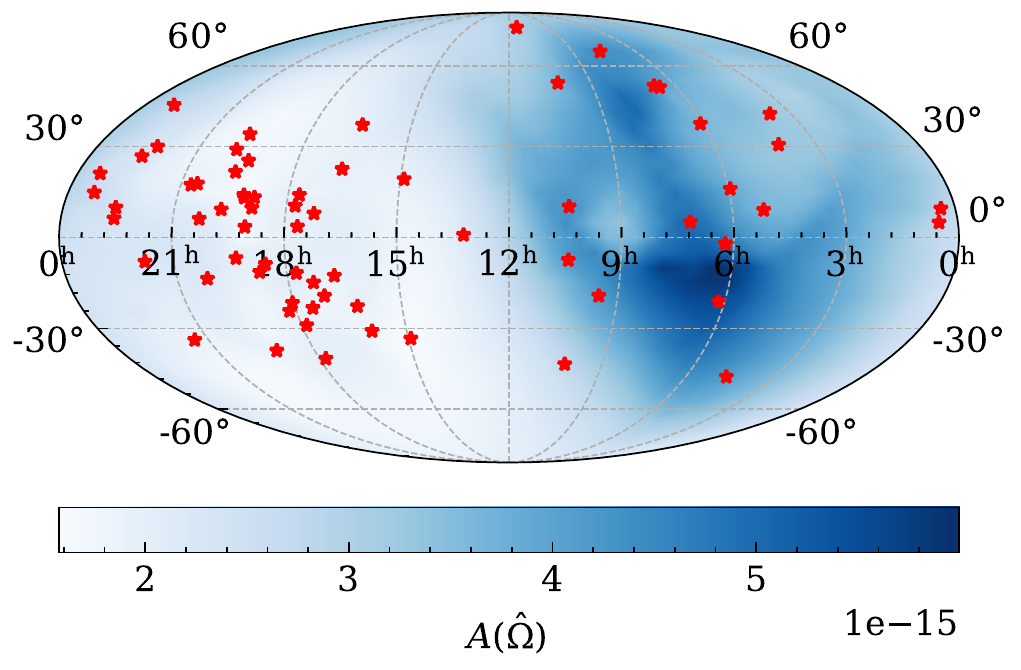}}
            \caption{Power measured at each pixel using the Bayesian radiometer pixel basis, in units of characteristic strain, with the red stars showing the position of pulsars in the PTA. This map was constructed with $N_{\rm side} = 8$, resulting in 768 pixels on the sky. Since the prior on the GWB amplitude was log-uniform between $10^{-18}$ and $10^{-14}$, the power was also constrained to be positive across the sky. The sky map is qualitatively consistent with the map derived using the frequentist radiometer pixel basis analysis.}
            \label{fig:bayes_radio_pixel}
        \end{figure}

        We also show the results from a Bayesian radiometer pixel analysis in \autoref{fig:bayes_radio_pixel}, where we plot the median power in each pixel. Since the GWB strain amplitude priors in the Bayesian analysis are log-uniform between $10^{-18}-10^{-14}$, we intrinsically restrict the power to be positive across all sky. Even though we observe some pixels with $A(\hat{\Omega}) \geq 5 \times 10^{-15}$, the odds ratios imply the data prefer an isotropic all-sky GWB over GWs originating from these pixels.
        
\section{Discussion} \label{sec:discussion}

    In this analysis, we search for anisotropy in the NANOGrav 15 yr dataset using the full set of correlation information. We do not find significant evidence for the presence of either power-law or frequency-resolved anisotropy.

    However, we observe features in the reconstructed sky maps that could be an early indication of anisotropy in the GWB. As shown in \autoref{fig:pl_maps}, we recover larger limits on deviations from isotropy at RA $\approx$3$^{\rm h}$ in the maps derived using the Bayesian square-root spherical harmonic analyses. In the frequency-resolved Bayesian anisotropy analysis, this same approximate feature appears in all of the lowest five frequency bins. This is likely indicative of the power-law template (restricted to the lowest five frequency bins) producing a representation of anisotropy that is averaged across the frequency bins used in the analysis. There is also excess power in this location in both the frequentist and Bayesian radiometer pixel analyses, though the lack of a significant sharp and localized (i.e., pixel-scale) feature implies that this feature may not be due to a single point GW source like a SMBHB. We also note that the upper limits derived here on the anisotropy in different directions on the sky are consistent with the limits set in a directed search for GWs from individual SMBHB systems \citep{ng12p5_cw}.

    We also show that there is no significant evidence (\autoref{fig:bayesian_bin2}) for anisotropy in the second frequency bin, where \citetalias{aaa+23} reported excess power with ``monopolar''\footnote{Note that monopolar correlations here refer to a correlation signature described by a constant offset as in \citetalias{aaa+23}, and not the ``monopole" as referred to in the spherical harmonic basis where it represents the isotropic component of the signal.} correlations. The reconstructed sky map (top right panel of \autoref{fig:fs_pow_map}) also does not show features that are significantly different from the maps produced for other frequency bins. Thus, we are unable to confirm if the ``monopolar'' correlation signature observed in \citetalias{aaa+23} is due to the presence of anisotropy in this frequency bin.
    
    While analytic \citep{ming_pta_anis, hotinli, spk_model}, semi-analytic \citep{localSMBHBs2017}, and simulation-based \citep{taylor_pta_anis, bumpy_bkgrnd} estimates for SMBHB-produced anisotropy have been proposed, they are all dependent on different model choices for the populations. However, we can use the GWB parameters measured in \citetalias{aaa+23} to make estimates of the expected level of anisotropy for SMBHB systems that are distributed randomly on the sky. The random distribution of SMBHB systems on the sky implicitly assumes that the large-scale structure is isotropic at these distances, implying that the estimates on anisotropy produced here will be pessimistic in nature. However, these estimates can serve as targets for PTAs to achieve in forthcoming datasets, by growing the array and increasing timing precision.

    To calculate the estimates conditioned on the GWB parameters in \citetalias{aaa+23}, we use SMBHB populations generated with \texttt{holodeck} \citep{holodeck}, following semi-analytic prescriptions for galaxy stellar mass function (GSMF), galaxy pair fraction, galaxy merger time, and black hole---bulge mass relations. These populations are then evolved in time using a self-consistent binary evolution model to produce an expectation value for the number of binaries of each SMBHB parameter (see \citet{holodeck} and \citet{aaa+23_astro}, for details). We generate individual universe realizations by drawing randomly from a Poisson distribution around this expectation value. To model the anisotropy of populations consistent with current GWB measurements, we select the 100 samples that best match the 15 yr characteristic strain amplitude measurement of $h_c(0.10 \mathrm{yr})=11.2\times 10^{-15}$ at $f=0.10\mathrm{yr^{-1}}$ \citep{aaa+23} from a set of 1000 samples of varying GSMF, BH-bulge mass relation, and hardening time parameters.

    Because the loudest single sources determine the level of anisotropy \citep{becsy_ss_2022}, one can treat all but the 2000 loudest as perfectly isotropic. 
    Thus for each realization, we select the 2000 loudest single sources in each frequency bin, place these single sources randomly on a HEALpix map of the sky with $N_{\rm side} = 32$ \citep{healpix}, and divide the characteristic strain of the background (all other sources) evenly among all pixels on the map. Finally, we calculate the spherical harmonics of these maps using the HEALPix \texttt{anafast} program with $\ell_\mathrm{max}=6$, to maintain consistency with the detection analysis presented in this work. 
    
\begin{figure}[htb]
        \centering
        \includegraphics[width=1.0\columnwidth]{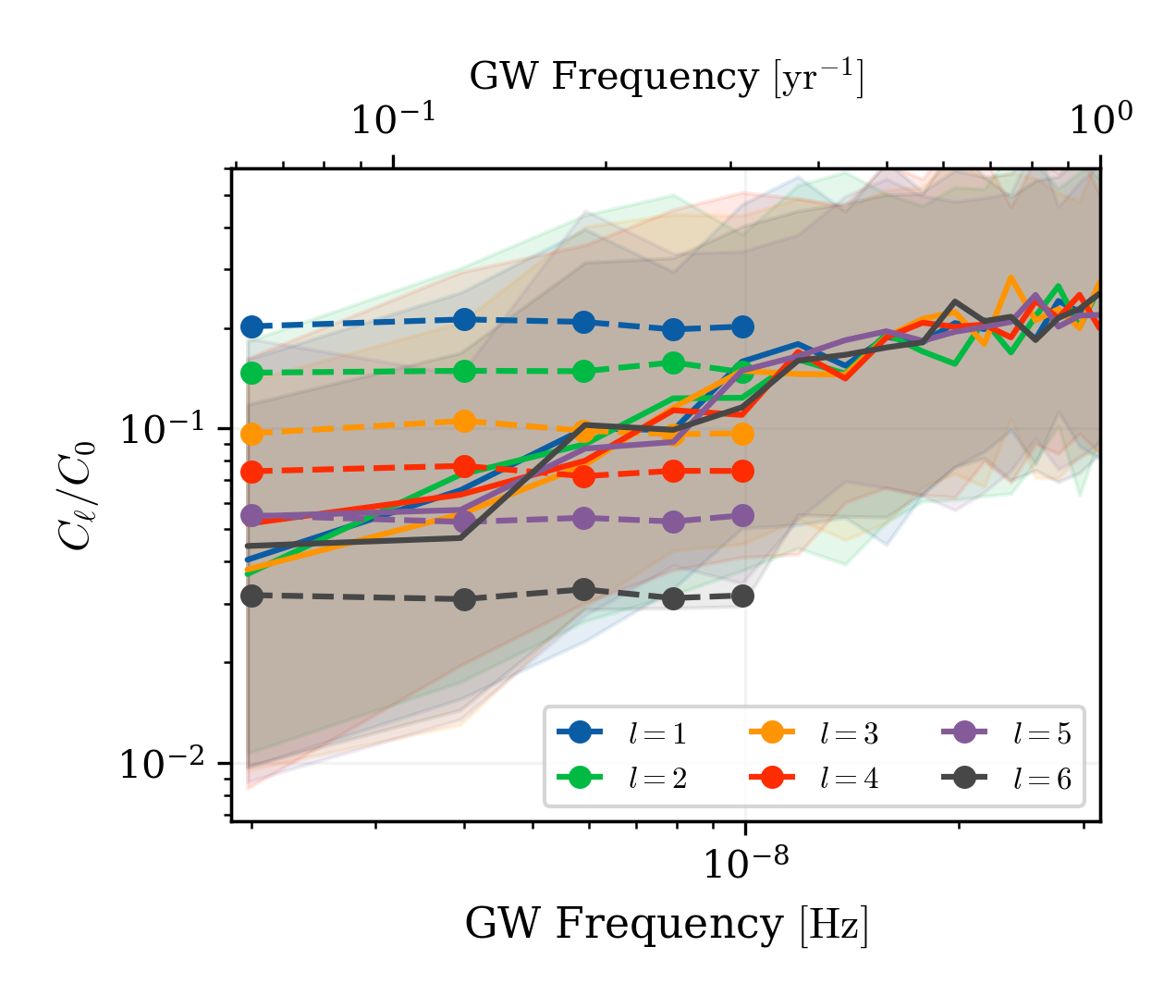}
        \caption{Normalized spherical harmonic coefficients $C_\ell/C_0$ of the gravitational wave sky as produced by simulated populations of SMBHBs, filtered by consistency with the 15 yr isotropic gravitational wave background estimation \citep{aaa+23}. The different colors correspond to individual harmonics from $\ell=1$ to $\ell=6$. The solid lines represent the median realization of the median samples and the shaded regions represent the 68\% confidence intervals across all samples' median realizations. The circles connected by dashed lines represent the Bayesian upper limits as in \autoref{fig:bayesian_cl}.}
        \label{fig:clc0_model}
    \end{figure}

    The normalized spherical harmonic coefficients $C_\ell /C_0$ of these samples are shown in \autoref{fig:clc0_model} with solid lines representing the median over samples and shaded regions being the 68\% confidence intervals. These samples' 68\% confidence interval span ~1 order of magnitude for all harmonics, but demonstrate a general power-law like increase in anisotropy with increasing frequency, their medians going from $C_\ell/C_0 \sim 4\times 10^{-2}$ at $f=0.10\ \mathrm{yr^{-1}}$ to $C_\ell/C_0 \sim 2 \times 10^{-1}$ at $f=1.0\ \mathrm{yr^{-1}}$. The results are indistinguishable between different harmonics of $\ell>0$, consistent with the analytic model analogous to large-scale structure shot noise in \citet{spk_model}. 

    For comparison, the Bayesian upper limits for spherical harmonics $\ell=1$ to $\ell=6$ from \autoref{fig:bayesian_cl} are plotted alongside the simulated SMBHB anisotropy in \autoref{fig:clc0_model}, as circles connected by dashed lines. Each upper limit falls within the 68\% confidence interval region for most frequencies, and the $\ell = 2$ to $\ell=5$ upper limits also intersect the median predictions. Unlike the simulated estimates of anisotropy which increase with frequency but are the same for all $\ell$, the upper limits are uniform across frequencies but decrease with $\ell$ by about 1 order of magnitude from $\ell=1$ to $\ell=6$. As such, we find these upper limits fall just below the 68\% confidence intervals for the lowest (non-zero) harmonics at low frequencies (2 nHz), and just below the simulated $C_\ell/C_0$ 68\% confidence intervals for high harmonic ($\ell=6$) at high frequencies (10 nHz). Thus, PTAs are likely to first detect small-scale anisotropy (larger values of $l$) before they detect large-scale anisotropy (smaller values of $l$) using the spherical harmonic basis.

\section{Conclusion and Future prospects} \label{sec:conclusion}

    We search for anisotropy in the NANOGrav 15 yr dataset and do not find significant evidence in favor of its presence in the GWB. As PTA datasets grow in time and add more pulsars to the array, the sensitivity of the detector to anisotropy in the GWB will increase \citep{bumpy_bkgrnd, pol_anisotropy} and allow us to conclusively determine if the features observed in the reconstructed GWB sky maps in this dataset are real. The International Pulsar Timing Array's \citep[IPTA,][]{IPTA} third data release (DR3), currently under preparation, will combine the NANOGrav 15 yr dataset with the latest datasets from the Eurpoean \citep{EPTA}, Indian \citep{InPTA}, and Parkes \citep{PPTA} PTAs. This combined dataset is projected to have approximately 80 pulsars (an increase of $\approx$20\% over the NANOGrav 15 yr dataset) and a maximum baseline of 24 yr (an increase of $\approx$60\% over the NANOGrav 15 yr dataset), and should consequently have better sensitivity to any anisotropy that might be present in the GWB. 

    As we head into this new era in nanohertz GW astronomy with PTA datasets growing their timing baselines and quantity of pulsars, new methods will need to be developed in order to efficiently search for anisotropy in these PTA datasets. The frequentist analyses implemented in this work are computationally efficient and have a small turnaround time for producing end-to-end results. However, the current implementation of the optimal statistic does not account for inter-pulsar-pair covariance \citep{pulsar_pair_correlation, cosmic_variance_2} and cosmic variance \citep{cosmic_variance_1}, and does not include information contained in the auto-correlations of the pulsar dataset. While the Bayesian analyses do not suffer from these drawbacks, they are computationally expensive and can take $\sim$weeks for the MCMC chains to burn-in and converge for the power-law anisotropy models, and even longer when searching for frequency-resolved anisotropy. It also takes a significant amount of time to scan across an $N_{\rm side} = 8$ sky in the Bayesian radiometer pixel analyses, and this time will increase with increasing values of $N_{\rm side}$. 
    
    One possible solution, currently under development, would be to formulate a method that can leverage Fourier basis coefficients used to model the GWB in the Bayesian analysis to directly calculate both the auto- and cross-correlations between the pulsars in the array. The frequentist framework developed in \citet{pol_anisotropy} and used in this work is agnostic to the method used to calculate the correlations and can be adapted to include the auto-correlation information. 
    Another solution would be to implement more sophisticated MCMC sampling techniques, such as reversible-jump MCMC (RJMCMC), which has already been implemented for single-source searches \citep{bayes_hopper}, and apply them to searches for anisotropy. RJMCMC, in particular, has the potential to significantly boost the efficiency of the Bayesian frequency-resolved and radiometer pixel searches since it has the capability to simultaneously search over multiple models (in this case frequencies or pixels respectively) as well as calculate the odds ratios between these models.
    Additionally, future mock data challenges like those carried out by the IPTA \citep{ipta_mdc_1, ipta_mdc_2} will be crucial in developing new methods as well as testing the efficacy of current methods to detect anisotropy in realistic PTA datasets.

    As the evidence for an isotropic GWB continues to grow with future datasets \citep{astro4cast}, it will become more prudent to search for the presence of anisotropy in the GWB. Since cosmological sources of a nanohertz GWB like cosmic strings are unlikely to produce an anisotropic GWB \citep{olmez_anisotropy}, detection of anisotropy will be an important piece of evidence in support of SMBHBs as the origin of the nanohertz GWB. A detection of pixel-scale anisotropy could provide the first indication of a single inspiraling SMBHB system which could then be subjected to targeted follow-up using both PTA GW and electromagnetic observatories. A detection of large scale anisotropy could indicate an over-density of SMBHB systems in a cluster environment, or might point to an extrinsic effect \citep{gw_kinematic_dipole}, such as the kinematic dipole observed with the cosmic microwave background \citep{cmb_kinematic_dipole} and predicted to be detectable with a GWB \citep{ gw_kinematic_dipole_4, gw_kinematic_dipole, gw_kinematic_dipole_2, gw_kinematic_dipole_3}. If we live in a Universe where both astrophysical (e.g., from SMBHBs) and cosmological (e.g., from cosmic strings) GWBs are present, we can leverage our knowledge of the different expected spatial (and spectral) distribution of these processes to disentangle them in the PTA datasets \citep{multiple_gwb_anis_1, multiple_gwb_anis_2, multiple_gwb_anis_5, multiple_gwb_anis_4, multiple_gwb_anis_3, multiple_anis_bkgnd, kaiser_source_confusion}. It will also be prudent to search for and possibly rule out anisotropy in the GWB before other interpretations of deviations away from the theoretical HD curve, such as beyond-GR effects \citep[e.g.,][]{ng_altpol}, are accepted.
    Thus, detection of anisotropy in the GWB is poised to be one of the next milestones in nanohertz GW astronomy, potentially leading us towards the detection of one or more isolated GW sources and allowing us to place constraints on beyond Standard Model physics.

\begin{acknowledgements}
    \emph{Author contributions.}
    \input{contrib}

    The NANOGrav collaboration receives support from National Science Foundation (NSF) Physics Frontiers Center award numbers 1430284 and 2020265, the Gordon and Betty Moore Foundation, NSF AccelNet award number 2114721, an NSERC Discovery Grant, and CIFAR. 
    The Arecibo Observatory is a facility of the NSF operated under cooperative agreement (AST-1744119) by the University of Central Florida (UCF) in alliance with Universidad Ana G. M{\'e}ndez (UAGM) and Yang Enterprises (YEI), Inc. The Green Bank Observatory is a facility of the NSF operated under cooperative agreement by Associated Universities, Inc. The National Radio Astronomy Observatory is a facility of the NSF operated under cooperative agreement by Associated Universities, Inc. 
    This work was conducted in part using the resources of the Advanced Computing Center for Research and Education (ACCRE) at Vanderbilt University, Nashville, TN.

    \input{acks15yranisotropy}

    \facilities{Arecibo, GBT, VLA}

    \software{\texttt{astropy} \citep{2022ApJ...935..167A}, \texttt{ENTERPRISE} \citep{enterprise}, \texttt{enterprise\_extensions} \citep{enterprise_ext}, \texttt{healpy} \citep{healpy}, \texttt{holodeck} \citep{holodeck}, \texttt{Jupyter} \citep{Kluyver2016jupyter}, \texttt{MAPS} \citep{pol_anisotropy}, \texttt{matplotlib} \citep{matplotlib}, \texttt{numpy} \citep{harris2020array}, \texttt{PTMCMC} \citep{ptmcmc}, \texttt{scipy} \citep{2020SciPy-NMeth}.}
    
    \vspace{12pt}
    
\end{acknowledgements}
    
\bibliography{bib.bib}
\bibliographystyle{aastex}

\end{document}

%% file: authors15yranisotropy.tex
\author[0000-0001-5134-3925]{Gabriella Agazie}
\affiliation{Center for Gravitation, Cosmology and Astrophysics, Department of Physics, University of Wisconsin-Milwaukee,\\ P.O. Box 413, Milwaukee, WI 53201, USA}
\author[0000-0002-8935-9882]{Akash Anumarlapudi}
\affiliation{Center for Gravitation, Cosmology and Astrophysics, Department of Physics, University of Wisconsin-Milwaukee,\\ P.O. Box 413, Milwaukee, WI 53201, USA}
\author[0000-0003-0638-3340]{Anne M. Archibald}
\affiliation{Newcastle University, NE1 7RU, UK}
\author{Zaven Arzoumanian}
\affiliation{X-Ray Astrophysics Laboratory, NASA Goddard Space Flight Center, Code 662, Greenbelt, MD 20771, USA}
\author[0000-0003-2745-753X]{Paul T. Baker}
\affiliation{Department of Physics and Astronomy, Widener University, One University Place, Chester, PA 19013, USA}
\author[0000-0003-0909-5563]{Bence B\'{e}csy}
\affiliation{Department of Physics, Oregon State University, Corvallis, OR 97331, USA}
\author[0000-0002-2183-1087]{Laura Blecha}
\affiliation{Physics Department, University of Florida, Gainesville, FL 32611, USA}
\author[0000-0001-6341-7178]{Adam Brazier}
\affiliation{Cornell Center for Astrophysics and Planetary Science and Department of Astronomy, Cornell University, Ithaca, NY 14853, USA}
\affiliation{Cornell Center for Advanced Computing, Cornell University, Ithaca, NY 14853, USA}
\author[0000-0003-3053-6538]{Paul R. Brook}
\affiliation{Institute for Gravitational Wave Astronomy and School of Physics and Astronomy, University of Birmingham, Edgbaston, Birmingham B15 2TT, UK}
\author[0000-0003-4052-7838]{Sarah Burke-Spolaor}
\affiliation{Department of Physics and Astronomy, West Virginia University, P.O. Box 6315, Morgantown, WV 26506, USA}
\affiliation{Center for Gravitational Waves and Cosmology, West Virginia University, Chestnut Ridge Research Building, Morgantown, WV 26505, USA}
\author[0000-0002-5557-4007]{J. Andrew Casey-Clyde}
\affiliation{Department of Physics, University of Connecticut, 196 Auditorium Road, U-3046, Storrs, CT 06269-3046, USA}
\author[0000-0003-3579-2522]{Maria Charisi}
\affiliation{Department of Physics and Astronomy, Vanderbilt University, 2301 Vanderbilt Place, Nashville, TN 37235, USA}
\author[0000-0002-2878-1502]{Shami Chatterjee}
\affiliation{Cornell Center for Astrophysics and Planetary Science and Department of Astronomy, Cornell University, Ithaca, NY 14853, USA}
\author[0000-0001-7587-5483]{Tyler Cohen}
\affiliation{Department of Physics, New Mexico Institute of Mining and Technology, 801 Leroy Place, Socorro, NM 87801, USA}
\author[0000-0002-4049-1882]{James M. Cordes}
\affiliation{Cornell Center for Astrophysics and Planetary Science and Department of Astronomy, Cornell University, Ithaca, NY 14853, USA}
\author[0000-0002-7435-0869]{Neil J. Cornish}
\affiliation{Department of Physics, Montana State University, Bozeman, MT 59717, USA}
\author[0000-0002-2578-0360]{Fronefield Crawford}
\affiliation{Department of Physics and Astronomy, Franklin \& Marshall College, P.O. Box 3003, Lancaster, PA 17604, USA}
\author[0000-0002-6039-692X]{H. Thankful Cromartie}
\altaffiliation{NASA Hubble Fellowship: Einstein Postdoctoral Fellow}
\affiliation{Cornell Center for Astrophysics and Planetary Science and Department of Astronomy, Cornell University, Ithaca, NY 14853, USA}
\author[0000-0002-1529-5169]{Kathryn Crowter}
\affiliation{Department of Physics and Astronomy, University of British Columbia, 6224 Agricultural Road, Vancouver, BC V6T 1Z1, Canada}
\author[0000-0002-2185-1790]{Megan E. DeCesar}
\affiliation{George Mason University, resident at the Naval Research Laboratory, Washington, DC 20375, USA}
\author[0000-0002-6664-965X]{Paul B. Demorest}
\affiliation{National Radio Astronomy Observatory, 1003 Lopezville Rd., Socorro, NM 87801, USA}
\author[0000-0001-8885-6388]{Timothy Dolch}
\affiliation{Department of Physics, Hillsdale College, 33 E. College Street, Hillsdale, MI 49242, USA}
\affiliation{Eureka Scientific, 2452 Delmer Street, Suite 100, Oakland, CA 94602-3017, USA}
\author{Brendan Drachler}
\affiliation{School of Physics and Astronomy, Rochester Institute of Technology, Rochester, NY 14623, USA}
\affiliation{Laboratory for Multiwavelength Astrophysics, Rochester Institute of Technology, Rochester, NY 14623, USA}
\author[0000-0001-7828-7708]{Elizabeth C. Ferrara}
\affiliation{Department of Astronomy, University of Maryland, College Park, MD 20742}
\affiliation{Center for Research and Exploration in Space Science and Technology, NASA/GSFC, Greenbelt, MD 20771}
\affiliation{NASA Goddard Space Flight Center, Greenbelt, MD 20771, USA}
\author[0000-0001-5645-5336]{William Fiore}
\affiliation{Department of Physics and Astronomy, West Virginia University, P.O. Box 6315, Morgantown, WV 26506, USA}
\affiliation{Center for Gravitational Waves and Cosmology, West Virginia University, Chestnut Ridge Research Building, Morgantown, WV 26505, USA}
\author[0000-0001-8384-5049]{Emmanuel Fonseca}
\affiliation{Department of Physics and Astronomy, West Virginia University, P.O. Box 6315, Morgantown, WV 26506, USA}
\affiliation{Center for Gravitational Waves and Cosmology, West Virginia University, Chestnut Ridge Research Building, Morgantown, WV 26505, USA}
\author[0000-0001-7624-4616]{Gabriel E. Freedman}
\affiliation{Center for Gravitation, Cosmology and Astrophysics, Department of Physics, University of Wisconsin-Milwaukee,\\ P.O. Box 413, Milwaukee, WI 53201, USA}
\author[0000-0002-8857-613X]{Emiko Gardiner}
\affiliation{Department of Astronomy, University of California, Berkeley, 501 Campbell Hall \#3411, Berkeley, CA 94720, USA}
\author[0000-0001-6166-9646]{Nate Garver-Daniels}
\affiliation{Department of Physics and Astronomy, West Virginia University, P.O. Box 6315, Morgantown, WV 26506, USA}
\affiliation{Center for Gravitational Waves and Cosmology, West Virginia University, Chestnut Ridge Research Building, Morgantown, WV 26505, USA}
\author[0000-0001-8158-683X]{Peter A. Gentile}
\affiliation{Department of Physics and Astronomy, West Virginia University, P.O. Box 6315, Morgantown, WV 26506, USA}
\affiliation{Center for Gravitational Waves and Cosmology, West Virginia University, Chestnut Ridge Research Building, Morgantown, WV 26505, USA}
\author[0000-0003-4090-9780]{Joseph Glaser}
\affiliation{Department of Physics and Astronomy, West Virginia University, P.O. Box 6315, Morgantown, WV 26506, USA}
\affiliation{Center for Gravitational Waves and Cosmology, West Virginia University, Chestnut Ridge Research Building, Morgantown, WV 26505, USA}
\author[0000-0003-1884-348X]{Deborah C. Good}
\affiliation{Department of Physics, University of Connecticut, 196 Auditorium Road, U-3046, Storrs, CT 06269-3046, USA}
\affiliation{Center for Computational Astrophysics, Flatiron Institute, 162 5th Avenue, New York, NY 10010, USA}
\author[0000-0002-1146-0198]{Kayhan G\"{u}ltekin}
\affiliation{Department of Astronomy and Astrophysics, University of Michigan, Ann Arbor, MI 48109, USA}
\author[0000-0003-2742-3321]{Jeffrey S. Hazboun}
\affiliation{Department of Physics, Oregon State University, Corvallis, OR 97331, USA}
\author[0000-0003-1082-2342]{Ross J. Jennings}
\altaffiliation{NANOGrav Physics Frontiers Center Postdoctoral Fellow}
\affiliation{Department of Physics and Astronomy, West Virginia University, P.O. Box 6315, Morgantown, WV 26506, USA}
\affiliation{Center for Gravitational Waves and Cosmology, West Virginia University, Chestnut Ridge Research Building, Morgantown, WV 26505, USA}
\author[0000-0002-7445-8423]{Aaron D. Johnson}
\affiliation{Center for Gravitation, Cosmology and Astrophysics, Department of Physics, University of Wisconsin-Milwaukee,\\ P.O. Box 413, Milwaukee, WI 53201, USA}
\affiliation{Division of Physics, Mathematics, and Astronomy, California Institute of Technology, Pasadena, CA 91125, USA}
\author[0000-0001-6607-3710]{Megan L. Jones}
\affiliation{Center for Gravitation, Cosmology and Astrophysics, Department of Physics, University of Wisconsin-Milwaukee,\\ P.O. Box 413, Milwaukee, WI 53201, USA}
\author[0000-0002-3654-980X]{Andrew R. Kaiser}
\affiliation{Department of Physics and Astronomy, West Virginia University, P.O. Box 6315, Morgantown, WV 26506, USA}
\affiliation{Center for Gravitational Waves and Cosmology, West Virginia University, Chestnut Ridge Research Building, Morgantown, WV 26505, USA}
\author[0000-0001-6295-2881]{David L. Kaplan}
\affiliation{Center for Gravitation, Cosmology and Astrophysics, Department of Physics, University of Wisconsin-Milwaukee,\\ P.O. Box 413, Milwaukee, WI 53201, USA}
\author[0000-0002-6625-6450]{Luke Zoltan Kelley}
\affiliation{Department of Astronomy, University of California, Berkeley, 501 Campbell Hall \#3411, Berkeley, CA 94720, USA}
\author[0000-0002-0893-4073]{Matthew Kerr}
\affiliation{Space Science Division, Naval Research Laboratory, Washington, DC 20375-5352, USA}
\author[0000-0003-0123-7600]{Joey S. Key}
\affiliation{University of Washington Bothell, 18115 Campus Way NE, Bothell, WA 98011, USA}
\author[0000-0002-9197-7604]{Nima Laal}
\affiliation{Department of Physics, Oregon State University, Corvallis, OR 97331, USA}
\author[0000-0003-0721-651X]{Michael T. Lam}
\affiliation{School of Physics and Astronomy, Rochester Institute of Technology, Rochester, NY 14623, USA}
\affiliation{Laboratory for Multiwavelength Astrophysics, Rochester Institute of Technology, Rochester, NY 14623, USA}
\author[0000-0003-1096-4156]{William G. Lamb}
\affiliation{Department of Physics and Astronomy, Vanderbilt University, 2301 Vanderbilt Place, Nashville, TN 37235, USA}
\author{T. Joseph W. Lazio}
\affiliation{Jet Propulsion Laboratory, California Institute of Technology, 4800 Oak Grove Drive, Pasadena, CA 91109, USA}
\author[0000-0003-0771-6581]{Natalia Lewandowska}
\affiliation{Department of Physics, State University of New York at Oswego, Oswego, NY, 13126, USA}
\author[0000-0001-5766-4287]{Tingting Liu}
\affiliation{Department of Physics and Astronomy, West Virginia University, P.O. Box 6315, Morgantown, WV 26506, USA}
\affiliation{Center for Gravitational Waves and Cosmology, West Virginia University, Chestnut Ridge Research Building, Morgantown, WV 26505, USA}
\author[0000-0003-1301-966X]{Duncan R. Lorimer}
\affiliation{Department of Physics and Astronomy, West Virginia University, P.O. Box 6315, Morgantown, WV 26506, USA}
\affiliation{Center for Gravitational Waves and Cosmology, West Virginia University, Chestnut Ridge Research Building, Morgantown, WV 26505, USA}
\author[0000-0001-5373-5914]{Jing Luo}
\altaffiliation{Deceased}
\affiliation{Department of Astronomy \& Astrophysics, University of Toronto, 50 Saint George Street, Toronto, ON M5S 3H4, Canada}
\author[0000-0001-5229-7430]{Ryan S. Lynch}
\affiliation{Green Bank Observatory, P.O. Box 2, Green Bank, WV 24944, USA}
\author[0000-0002-4430-102X]{Chung-Pei Ma}
\affiliation{Department of Astronomy, University of California, Berkeley, 501 Campbell Hall \#3411, Berkeley, CA 94720, USA}
\affiliation{Department of Physics, University of California, Berkeley, CA 94720, USA}
\author[0000-0003-2285-0404]{Dustin R. Madison}
\affiliation{Department of Physics, University of the Pacific, 3601 Pacific Avenue, Stockton, CA 95211, USA}
\author[0000-0001-5481-7559]{Alexander McEwen}
\affiliation{Center for Gravitation, Cosmology and Astrophysics, Department of Physics, University of Wisconsin-Milwaukee,\\ P.O. Box 413, Milwaukee, WI 53201, USA}
\author[0000-0002-2885-8485]{James W. McKee}
\affiliation{E.A. Milne Centre for Astrophysics, University of Hull, Cottingham Road, Kingston-upon-Hull, HU6 7RX, UK}
\affiliation{Centre of Excellence for Data Science, Artificial Intelligence and Modelling (DAIM), University of Hull, Cottingham Road, Kingston-upon-Hull, HU6 7RX, UK}
\author[0000-0001-7697-7422]{Maura A. McLaughlin}
\affiliation{Department of Physics and Astronomy, West Virginia University, P.O. Box 6315, Morgantown, WV 26506, USA}
\affiliation{Center for Gravitational Waves and Cosmology, West Virginia University, Chestnut Ridge Research Building, Morgantown, WV 26505, USA}
\author[0000-0002-4642-1260]{Natasha McMann}
\affiliation{Department of Physics and Astronomy, Vanderbilt University, 2301 Vanderbilt Place, Nashville, TN 37235, USA}
\author[0000-0001-8845-1225]{Bradley W. Meyers}
\affiliation{Department of Physics and Astronomy, University of British Columbia, 6224 Agricultural Road, Vancouver, BC V6T 1Z1, Canada}
\affiliation{International Centre for Radio Astronomy Research, Curtin University, Bentley, WA 6102, Australia}
\author[0000-0002-4307-1322]{Chiara M. F. Mingarelli}
\affiliation{Center for Computational Astrophysics, Flatiron Institute, 162 5th Avenue, New York, NY 10010, USA}
\affiliation{Department of Physics, University of Connecticut, 196 Auditorium Road, U-3046, Storrs, CT 06269-3046, USA}
\affiliation{Department of Physics, Yale University, New Haven, CT 06520, USA}
\author[0000-0003-2898-5844]{Andrea Mitridate}
\affiliation{Deutsches Elektronen-Synchrotron DESY, Notkestr. 85, 22607 Hamburg, Germany}
\author[0000-0002-3616-5160]{Cherry Ng}
\affiliation{Dunlap Institute for Astronomy and Astrophysics, University of Toronto, 50 St. George St., Toronto, ON M5S 3H4, Canada}
\author[0000-0002-6709-2566]{David J. Nice}
\affiliation{Department of Physics, Lafayette College, Easton, PA 18042, USA}
\author[0000-0002-4941-5333]{Stella Koch Ocker}
\affiliation{Cornell Center for Astrophysics and Planetary Science and Department of Astronomy, Cornell University, Ithaca, NY 14853, USA}
\author[0000-0002-2027-3714]{Ken D. Olum}
\affiliation{Institute of Cosmology, Department of Physics and Astronomy, Tufts University, Medford, MA 02155, USA}
\author[0000-0001-5465-2889]{Timothy T. Pennucci}
\affiliation{Institute of Physics and Astronomy, E\"{o}tv\"{o}s Lor\'{a}nd University, P\'{a}zm\'{a}ny P. s. 1/A, 1117 Budapest, Hungary}
\author[0000-0002-8509-5947]{Benetge B. P. Perera}
\affiliation{Arecibo Observatory, HC3 Box 53995, Arecibo, PR 00612, USA}
\author[0000-0002-8826-1285]{Nihan S. Pol}
\affiliation{Department of Physics and Astronomy, Vanderbilt University, 2301 Vanderbilt Place, Nashville, TN 37235, USA}
\author[0000-0002-2074-4360]{Henri A. Radovan}
\affiliation{Department of Physics, University of Puerto Rico, Mayag\"{u}ez, PR 00681, USA}
\author[0000-0001-5799-9714]{Scott M. Ransom}
\affiliation{National Radio Astronomy Observatory, 520 Edgemont Road, Charlottesville, VA 22903, USA}
\author[0000-0002-5297-5278]{Paul S. Ray}
\affiliation{Space Science Division, Naval Research Laboratory, Washington, DC 20375-5352, USA}
\author[0000-0003-4915-3246]{Joseph D. Romano}
\affiliation{Department of Physics, Texas Tech University, Box 41051, Lubbock, TX 79409, USA}
\author[0009-0006-5476-3603]{Shashwat C. Sardesai}
\affiliation{Center for Gravitation, Cosmology and Astrophysics, Department of Physics, University of Wisconsin-Milwaukee,\\ P.O. Box 413, Milwaukee, WI 53201, USA}
\author[0000-0003-4391-936X]{Ann Schmiedekamp}
\affiliation{Department of Physics, Penn State Abington, Abington, PA 19001, USA}
\author[0000-0002-1283-2184]{Carl Schmiedekamp}
\affiliation{Department of Physics, Penn State Abington, Abington, PA 19001, USA}
\author[0000-0003-2807-6472]{Kai Schmitz}
\affiliation{Institute for Theoretical Physics, University of M\"{u}nster, 48149 M\"{u}nster, Germany}
\author[0000-0001-6425-7807]{Levi Schult}
\affiliation{Department of Physics and Astronomy, Vanderbilt University, 2301 Vanderbilt Place, Nashville, TN 37235, USA}
\author[0000-0002-7283-1124]{Brent J. Shapiro-Albert}
\affiliation{Department of Physics and Astronomy, West Virginia University, P.O. Box 6315, Morgantown, WV 26506, USA}
\affiliation{Center for Gravitational Waves and Cosmology, West Virginia University, Chestnut Ridge Research Building, Morgantown, WV 26505, USA}
\affiliation{Giant Army, 915A 17th Ave, Seattle WA 98122}
\author[0000-0002-7778-2990]{Xavier Siemens}
\affiliation{Department of Physics, Oregon State University, Corvallis, OR 97331, USA}
\affiliation{Center for Gravitation, Cosmology and Astrophysics, Department of Physics, University of Wisconsin-Milwaukee,\\ P.O. Box 413, Milwaukee, WI 53201, USA}
\author[0000-0003-1407-6607]{Joseph Simon}
\altaffiliation{NSF Astronomy and Astrophysics Postdoctoral Fellow}
\affiliation{Department of Astrophysical and Planetary Sciences, University of Colorado, Boulder, CO 80309, USA}
\author[0000-0002-1530-9778]{Magdalena S. Siwek}
\affiliation{Center for Astrophysics, Harvard University, 60 Garden St, Cambridge, MA 02138}
\author[0000-0001-9784-8670]{Ingrid H. Stairs}
\affiliation{Department of Physics and Astronomy, University of British Columbia, 6224 Agricultural Road, Vancouver, BC V6T 1Z1, Canada}
\author[0000-0002-1797-3277]{Daniel R. Stinebring}
\affiliation{Department of Physics and Astronomy, Oberlin College, Oberlin, OH 44074, USA}
\author[0000-0002-7261-594X]{Kevin Stovall}
\affiliation{National Radio Astronomy Observatory, 1003 Lopezville Rd., Socorro, NM 87801, USA}
\author[0000-0002-2820-0931]{Abhimanyu Susobhanan}
\affiliation{Center for Gravitation, Cosmology and Astrophysics, Department of Physics, University of Wisconsin-Milwaukee,\\ P.O. Box 413, Milwaukee, WI 53201, USA}
\author[0000-0002-1075-3837]{Joseph K. Swiggum}
\altaffiliation{NANOGrav Physics Frontiers Center Postdoctoral Fellow}
\affiliation{Department of Physics, Lafayette College, Easton, PA 18042, USA}
\author[0000-0003-0264-1453]{Stephen R. Taylor}
\affiliation{Department of Physics and Astronomy, Vanderbilt University, 2301 Vanderbilt Place, Nashville, TN 37235, USA}
\author[0000-0002-2451-7288]{Jacob E. Turner}
\affiliation{Department of Physics and Astronomy, West Virginia University, P.O. Box 6315, Morgantown, WV 26506, USA}
\affiliation{Center for Gravitational Waves and Cosmology, West Virginia University, Chestnut Ridge Research Building, Morgantown, WV 26505, USA}
\author[0000-0001-8800-0192]{Caner Unal}
\affiliation{Department of Physics, Ben-Gurion University of the Negev, Be'er Sheva 84105, Israel}
\affiliation{Feza Gursey Institute, Bogazici University, Kandilli, 34684, Istanbul, Turkey}
\author[0000-0002-4162-0033]{Michele Vallisneri}
\affiliation{Jet Propulsion Laboratory, California Institute of Technology, 4800 Oak Grove Drive, Pasadena, CA 91109, USA}
\affiliation{Division of Physics, Mathematics, and Astronomy, California Institute of Technology, Pasadena, CA 91125, USA}
\author[0000-0003-4700-9072]{Sarah J. Vigeland}
\affiliation{Center for Gravitation, Cosmology and Astrophysics, Department of Physics, University of Wisconsin-Milwaukee,\\ P.O. Box 413, Milwaukee, WI 53201, USA}
\author[0000-0001-9678-0299]{Haley M. Wahl}
\affiliation{Department of Physics and Astronomy, West Virginia University, P.O. Box 6315, Morgantown, WV 26506, USA}
\affiliation{Center for Gravitational Waves and Cosmology, West Virginia University, Chestnut Ridge Research Building, Morgantown, WV 26505, USA}
\author[0000-0002-6020-9274]{Caitlin A. Witt}
\affiliation{Center for Interdisciplinary Exploration and Research in Astrophysics (CIERA), Northwestern University, Evanston, IL 60208}
\affiliation{Adler Planetarium, 1300 S. DuSable Lake Shore Dr., Chicago, IL 60605, USA}
\author[0000-0002-0883-0688]{Olivia Young}
\affiliation{School of Physics and Astronomy, Rochester Institute of Technology, Rochester, NY 14623, USA}
\affiliation{Laboratory for Multiwavelength Astrophysics, Rochester Institute of Technology, Rochester, NY 14623, USA}

%% file: contrib.tex
An alphabetical-order author list was used for this paper in recognition of the fact that a large, decade timescale project such as NANOGrav is necessarily the result of the work of many people. All authors contributed to the activities of the NANOGrav collaboration leading to the work presented here, and reviewed the manuscript, text, and figures prior to the paper's submission. 
Additional specific contributions to this paper are as follows.
%
N.S.P., S.R.T., and J.D.R. proposed and planned this analysis. N.S.P. co-ordinated the execution of the analysis and writing of this paper. N.S.P., S.R.T., E.C.G., L.Z.K., J.D.R., K.O., P.M., N.L. contributed text to the paper. N.S.P. led the frequentist analyses, N.L. and L.S. led the Bayesian analyses, and E.C.G. led the \textsc{holodeck}-based astrophysical simulation analyses, all of which had input from S.R.T, J.D.R, L.Z.K., K.O., and P.M.
The NANOGrav 15-year data set was developed by G.A., A.A., A.M.A., Z.A., P.T.B., P.R.B., H.T.C., K.C., M.E.D., P.B.D., T.D., E.C.F., W.F., E.F., G.E.F., N.G., P.A.G., J.G., D.C.G., J.S.H., R.J.J., M.L.J., D.L.K., M.K., M.T.L., D.R.L., J.L., R.S.L., A.M., M.A.M., N.M., B.W.M., C.N., D.J.N., T.T.P., B.B.P.P., N.S.P., H.A.R., S.M.R., P.S.R., A.S., C.S., B.J.S., I.H.S., K.S., A.S., J.K.S., and H.M.W. through a combination of 
observations, arrival time calculations, data checks and refinements, 
and timing model development and analysis; 
additional specific contributions to the data set are summarized in \citetalias{aaa+23_dataset}.

%% file: acks15yranisotropy.tex
L.B. acknowledges support from the National Science Foundation under award AST-1909933 and from the Research Corporation for Science Advancement under Cottrell Scholar Award No. 27553.
P.R.B. is supported by the Science and Technology Facilities Council, grant number ST/W000946/1.
S.B. gratefully acknowledges the support of a Sloan Fellowship, and the support of NSF under award \#1815664.
M.C. and S.R.T. acknowledge support from NSF AST-2007993.
M.C. and N.S.P. were supported by the Vanderbilt Initiative in Data Intensive Astrophysics (VIDA) Fellowship.
Support for this work was provided by the NSF through the Grote Reber Fellowship Program administered by Associated Universities, Inc./National Radio Astronomy Observatory.
Support for H.T.C. is provided by NASA through the NASA Hubble Fellowship Program grant \#HST-HF2-51453.001 awarded by the Space Telescope Science Institute, which is operated by the Association of Universities for Research in Astronomy, Inc., for NASA, under contract NAS5-26555.
Pulsar research at UBC is supported by an NSERC Discovery Grant and by CIFAR.
K.C. is supported by a UBC Four Year Fellowship (6456).
M.E.D. acknowledges support from the Naval Research Laboratory by NASA under contract S-15633Y.
T.D. and M.T.L. are supported by an NSF Astronomy and Astrophysics Grant (AAG) award number 2009468.
E.C.F. is supported by NASA under award number 80GSFC21M0002.
G.E.F., S.C.S., and S.J.V. are supported by NSF award PHY-2011772.
The Flatiron Institute is supported by the Simons Foundation.
A.D.J. and M.V. acknowledge support from the Caltech and Jet Propulsion Laboratory President's and Director's Research and Development Fund.
A.D.J. acknowledges support from the Sloan Foundation.
The work of N.La. and X.S. is partly supported by the George and Hannah Bolinger Memorial Fund in the College of Science at Oregon State University.
N.La. acknowledges the support from Larry W. Martin and Joyce B. O'Neill Endowed Fellowship in the College of Science at Oregon State University.
Part of this research was carried out at the Jet Propulsion Laboratory, California Institute of Technology, under a contract with the National Aeronautics and Space Administration (80NM0018D0004).
D.R.L. and M.A.M. are supported by NSF \#1458952.
M.A.M. is supported by NSF \#2009425.
C.M.F.M. was supported in part by the National Science Foundation under Grants No. NSF PHY-1748958 and AST-2106552.
A.Mi. is supported by the Deutsche Forschungsgemeinschaft under Germany's Excellence Strategy - EXC 2121 Quantum Universe - 390833306.
The Dunlap Institute is funded by an endowment established by the David Dunlap family and the University of Toronto.
K.D.O. was supported in part by NSF Grant No. 2207267.
T.T.P. acknowledges support from the Extragalactic Astrophysics Research Group at E\"{o}tv\"{o}s Lor\'{a}nd University, funded by the E\"{o}tv\"{o}s Lor\'{a}nd Research Network (ELKH), which was used during the development of this research.
S.M.R. and I.H.S. are CIFAR Fellows.
Portions of this work performed at NRL were supported by ONR 6.1 basic research funding.
J.D.R. also acknowledges support from start-up funds from Texas Tech University.
J.S. is supported by an NSF Astronomy and Astrophysics Postdoctoral Fellowship under award AST-2202388, and acknowledges previous support by the NSF under award 1847938.
S.R.T. acknowledges support from an NSF CAREER award \#2146016.
C.U. acknowledges support from BGU (Kreitman fellowship), and the Council for Higher Education and Israel Academy of Sciences and Humanities (Excellence fellowship).
C.A.W. acknowledges support from CIERA, the Adler Planetarium, and the Brinson Foundation through a CIERA-Adler postdoctoral fellowship.
O.Y. is supported by the National Science Foundation Graduate Research Fellowship under Grant No. DGE-2139292.